\documentclass[reprint,superscriptaddress,amsmath,amssymb,aps,pra,floatfix]{revtex4-2}
\usepackage{multirow}
\usepackage{chemfig}
\usepackage{xcolor}
\usepackage{graphicx}
\usepackage{dcolumn}
\usepackage{bm}
\usepackage{hyperref}
\hypersetup{colorlinks=true, urlcolor=blue, linkcolor=blue, citecolor=blue} 
\usepackage[T1]{fontenc}
\usepackage{tikz}

\definecolor{matlabblue}{rgb}{0,0.45,0.74}

\newcommand{\filleddot}[2][black]{%
\tikz \fill[#1] (0,0) circle (#2);%
}

\newcommand{\filledStarOfDavid}[2][black]{%
\begin{tikzpicture}[baseline=-0.6ex, scale=#2]
  \fill[#1] (90:1) -- (210:1) -- (330:1) -- cycle;
  \fill[#1] (270:1) -- (30:1) -- (150:1) -- cycle;
\end{tikzpicture}%
}

\newcommand{\filleddowntriangle}[2][black]{%
\begin{tikzpicture}[baseline=-0.6ex, scale=#2]
  \fill[#1] (270:1) -- (30:1) -- (150:1) -- cycle;
\end{tikzpicture}%
}

\newcommand{\filledtriangle}[2][black]{%
\begin{tikzpicture}[baseline=-0.6ex, scale=#2]
  \fill[#1] (90:1) -- (210:1) -- (330:1) -- cycle;
\end{tikzpicture}%
}

\begin{document}

\title{Study of low-energy electron-induced dissociation of 1-Propanol}

\author{Soumya Ghosh}
\email{evan.soumya@gmail.com}
\affiliation{Department of Physical Sciences, Indian Institute of Science Education and Research Kolkata, Mohanpur-741246, India}

\author{Dipayan Chakraborty}
\email{physics.dipayan@gmail.com}
\affiliation{Department of Physical Sciences, Indian Institute of Science Education and Research Kolkata, Mohanpur-741246, India}

\author{Anirban Paul}
\thanks{Present address: J. Heyrovský Institute of Physical Chemistry, Czech Academy of Sciences, Dolejškova 3, 182 23 Prague, Czech Republic}
\email{anirban.paul1995@gmail.com}
\affiliation{Department of Physical Sciences, Indian Institute of Science Education and Research Kolkata, Mohanpur-741246, India}

\author{Dhananjay Nandi}
\email{dhananjay@iiserkol.ac.in}
\affiliation{Department of Physical Sciences, Indian Institute of Science Education and Research Kolkata, Mohanpur-741246, India}
\affiliation{Center for Atomic, Molecular and Optical Sciences and Technologies, Joint initiative of IIT Tirupati and IISER Tirupati, Yerpedu, 517619, Andhra Pradesh, India}

\begin{abstract}
The fragmentation of 1-propanol resulting from dissociative electron attachment has been explored across an energy range of 3.5 to 16 eV. Four distinct ion species are identified: $\text{H}^{-}$, $\text{O}^{-}$, $\text{OH}^{-}$, and $\text{C}_{3}\text{H}_{7}\text{O}^{-}$. The $\text{OH}^{-}$ ion exhibited a prominent peak near 8.7 eV, along with a small hump near 5.6 eV. Complementary channels led to the formation of the $\text{H}^{-}$ and $\text{C}_{3}\text{H}_{7}\text{O}^{-}$ ions. Both these two ions exhibit a sharp peak near 6 eV and broad overlapping resonances between 7 to 12 eV. The observed ion yields of distinct dissociation fragments in this study, when compared with those from previously studied alcohols, suggest site-specific fragmentation of alcohols during dissociative electron attachment. To gain a deeper understanding of the dissociation pathways, Density Functional Theory~(DFT) calculations were conducted, revealing the threshold energies for each channel. These threshold energies aligned well with the experimental uncertainties.
\end{abstract}

\maketitle

\section{\label{sec:introduction}Introduction}

Dissociative electron attachment (DEA) is one of several key mechanisms by which low-energy electrons can break molecules apart, generating reactive anions and radicals~\cite{Fabrikant_Review,ingolfsson_low-energy_2019}. In DEA, a transient negative ion (TNI) state forms when an electron attaches to a molecule. If the resonance persists for a sufficient duration before autodetachment occurs, it can decay by breaking a bond~\cite{Illenberger_book}. Therefore, the cross section of a DEA resonance is governed by two key factors: the probability of electron attachment to the neutral molecule and the survival probability of the resulting transient negative ion. Among these resonances, Feshbach resonances are special electronic states within TNIs that cannot autoionize through a simple single-electron transition~\cite{Schulz}. These core-excited states are known to be critically important in DEA due to their larger cross-section, but they pose major challenges for even advanced theoretical approaches, as accurately describing their electronic correlation is essential to predict their potential energy surfaces and lifetimes~\cite{Chakraborty_Ammonia,Adaniya_Water,Chakraborty:2023}. Experimental studies measuring resonance energies and DEA ion yields have offered valuable information about the behavior and characteristics of Feshbach resonances. Feshbach resonances are prevalent in alcohols and have been documented across a diverse spectrum of energies, extending as low as a few electron volts below the ionization potential~\cite{Chakraborty:2023,Prabhudesai:2005,Ibanescu:2007,Paul:2023}.

The total cross-section for electron scattering by primary alcohols is essential for understanding molecular reactivity in collision processes and has applications in optimizing alcohol combustion in internal combustion engines. This optimization can enhance performance and promote the use of renewable sources. Ethanol, the most widely used alternative fuel, is often blended with fossil fuels to reduce greenhouse gas emissions. It has a higher octane rating (99.5 AKI) than gasoline (85-96 AKI), enhancing engine performance, but its lower energy density (20 MJ/l) compared to gasoline (33 MJ/l) results in reduced fuel economy. This suggests the potential of using larger alcohols like propanol, which has a higher octane rating (108 AKI) and better energy density (24 MJ/l), as alternative fuels. DEA studies contribute to the comprehension of electron-induced degradation of propanol in high-energy combustion, plasma-assisted ignition, and fuel injectors subjected to ionizing environments. The cross-sectional studies are essential in this regard and can justify the ongoing intensive research.

The two simplest alcohols, methanol and ethanol, have been extensively investigated, both theoretically and experimentally, over the past two decades~\cite{NIXON:2016, C_Szmytkowski_1995, M_Vinodkumar_2008, Silva_2010, TAN:2011, Vinodkumar:2013, Lee:2012, Bouchiha_2007, Khakoo_methanol:2008, Sugohara:2011, Brunger_2017, djuric:1990, Rejoub:2003, Srivastava:1996, Hudson:2003, deutsch:1998, PAL:2004, VINODKUMAR:2011, trepka:1963, Kuhn:1988, Curtis:1992, Prabhudesai:2005, Skalicky_2004, Ibanescu:2007, Prabhudesai_2008_jcp, Wang:2015, orzol:2007, Paul:2023, Chakraborty:2023}. Relatively few data available for $1-$propanol~\cite{djuric:1990,Rejoub:2003,Hudson:2003,VINODKUMAR:2011,Khakoo:2008,Williams:1968,Bull:2012,TAKEUCHI:1985,PIRES:2017,Silva:2017}. The theoretical and experimental differential cross section for elastic scattering~\cite{Khakoo:2008}, total ionization cross section~\cite{Hudson:2003,Silva:2017}, partial ionization cross section~\cite{Rejoub:2003,PIRES:2017}. Takeuchi {\em et~al.}~\cite{TAKEUCHI:1985} had discussed the fragmentation method and potential energy curves. DEA to $1$-propanol has only been studied by Ibănescu~\cite{IbanescuBC:phd}. The authors' compared the photoelectron spectra and DEA spectra, noting a shift of $-4.2$ eV. Four different masses were identified: $59$ amu $\left(\text{C}_{3}\text{H}_{7}\text{O}^{-}\right)$, $57$ amu $\left(\text{C}_{3}\text{H}_{5}\text{O}^{-}\right)$, $17$ amu $\left(\text{OH}^{-}\right)$, and $1$ amu $\left(\text{H}^{-}\right)$. Some resonances were discussed briefly, and the excitation functions of H$^-$, OH$^-$, and H-loss anions were provided. To date, there have been very limited studies on DEA to 1-propanol. In this article, we examine the resonances of different fragment anions resulting from DEA to $1$-propanol.

 \section{\label{sec2:experimental}Experimental and computational method}

In this study, a segmented time-of-flight mass spectrometer is utilized. The original experimental setup and methodology have been described in detail in a previous publication~\cite{Chakraborty:2018}. Later, the spectrometer was replaced with a longer version with additional electrodes, which resulted in better mass resolution with a relatively low voltage repellent pulse. This modification allows the detection of H$^-$ ions. The detail of this modified spectrometer is reported in a separate work~\cite{Paul:2023}; therefore, only a brief overview is provided here.

A pulsed, magnetically collimated electron beam is produced using a custom-built electron gun, where thermionic emission is initiated by resistively heating a tungsten filament. The emitted electrons are focused through a set of electrodes maintained at specific bias voltages. To control the electron pulse, one electrode is held at a negative potential to suppress electron flow, followed by a positive voltage pulse that releases the electrons, thereby generating a well-defined pulsed beam. The electron gun is housed within a pair of Helmholtz coils that produce a uniform magnetic field along the beam axis, ensuring collimation of the electrons as they exit the source. The beam operates at a repetition rate of 10 kHz with a pulse width of 200 ns and is directed orthogonally to the molecular beam. A Faraday cup, positioned just downstream of the interaction region along the electron beam axis, is used to collect the electrons and measure the time-averaged beam current.

The continuous effusive molecular beam is introduced via a needle with a 1mm opening mounted inside the repeller plate, electrically isolated from it. The spectrometer is aligned along the molecular beam axis, which is orthogonal to the electron beam. Within the interaction region, electrons and molecules collide between the repeller and puller plates, leading to the formation of negative ions in DEA. The resulting negative ions are extracted and guided toward a micro channel plate (MCP) detector using a combination of segmented and linear ToF spectrometers. The spectrometer consists of a repeller plate, a puller plate, three lens electrodes arranged in an einzel lens configuration, a long field-free drift tube, and an additional set of einzel lenses positioned at the end of the flight tube to enhance ion trajectory control. The attractor plate is constructed from wire mesh with approximately 90\% ion transmission efficiency, allowing most ions to pass through while minimizing electric field penetration. A similar mesh is installed at the terminal end of the final electrode, immediately before the MCP detector, to shield the spectrometer from stray electric fields originating from the detector. The MCP detector is arranged in a chevron configuration to suppress ion feedback. Behind the MCPs, a Faraday cup—biased at a higher potential—collects the ion-induced charge cloud. This charge is coupled to a capacitor-decoupler circuit that captures the resulting voltage fluctuations or the electrical signals. These signals are then amplified by using a fast amplifier and then passed to a Constant Fraction Discriminator (CFD), which converts the analog signal into a standardized digital pulse. The CFD output serves as the stop signal for a Nuclear Instrumentation Module (NIM)-based Time-to-Amplitude Converter (TAC), while the start signal is synchronized with the electron beam pulse using a master pulse generator. The time difference between these start and stop signals yields the ion's time-of-flight. This signal is recorded by a Multichannel Analyzer (MCA) and processed using MAESTRO software. The ion ToF is proportional to the square root of the mass-to-charge ratio $\left(\text{T}\propto \sqrt{m/q}\right)$, where $m$ is the mass of the ions and $q$ is the charge. A specific mass-to-charge ratio ($m/q$) is selected and scanned across the entire electron energy range to measure the yield of the corresponding ions.

The mass calibration was conducted utilizing anionic fragments generated from DEA to SO$_2$ at an electron energy of 4.5 eV. The electron energy calibration was performed by utilizing the resonant peaks of O$^-$ ion yields from DEA to CO$_2$ at 4 eV and O$_2$ at 6.5 eV~\cite{Rapp:2004}.

We have performed the quantum chemical calculations using GAUSSIAN 16 software to determine the threshold energies of a few selected dissociation channels producing fragment anions~\cite{g16}. We have calculated the zero-point energy corrected electronic energies of each of the neutral and anionic fragments in their geometrically optimized structure. We used the B3LYP density functional theory (DFT) method~\cite{Becke_1993} with aug-cc-pVTZ basis sets~\cite{Dunning:1989} for that purpose. We have calculated the threshold energies of different dissociation channels as the differences between the total energies of the products (neutral and anionic fragments) and the target molecule (1-Propanol).

\section{\label{sec3:results}Result and discussion}

\begin{figure}[ht!]
\centering
\includegraphics[scale=0.47]{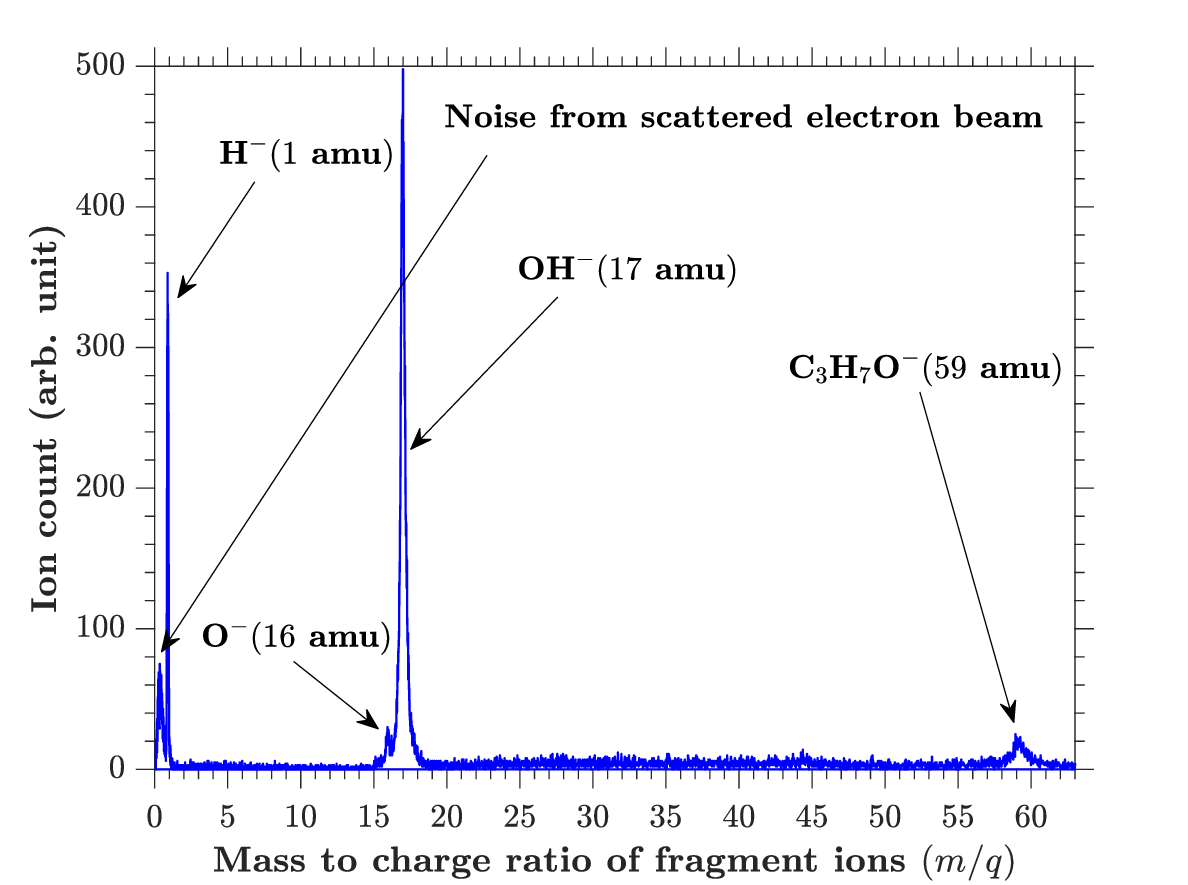}
\caption{\label{fig:mass_12v}Mass spectra recorded following the interaction of 10 eV electrons with 1-propanol reveal four distinct ion masses: H$^-$ (1 amu), O$^-$ (16 amu), OH$^-$ (17 amu), and C$_3$H$_7$O$^-$ (59 amu). The collection efficiency of H$^-$ ions is limited, as reported in Ref.~\cite{Paul:2023}.}
\end{figure}

The mass spectra of the fragment anions around $10$~eV electron energy is shown in Fig.~\ref{fig:mass_12v}. The TNI created in electron collision with $1-$propanol, gets dissociated in four different dissociation channels.
\begin{eqnarray*}
\label{eq1}
\text{CH}_3\text{CH}_2\text{CH}_2\text{OH}+ e^{-} &&\longrightarrow \left(\text{CH}_3\text{CH}_2\text{CH}_2\text{OH}\right)^{-*}\nonumber \\
\left(\text{CH}_3\text{CH}_2\text{CH}_2\text{OH}\right)^{-*} &&\longrightarrow
\begin{cases}
\text{H}^{-}+\text{Neutral(s)}\nonumber \\
\text{O}^{-}+\text{Neutral(s)}\nonumber \\
\text{OH}^{-}+\text{Neutral(s)}\nonumber \\
\text{C}_{3}\text{H}_{7}\text{O}^{-}+\text{Neutral(s)}\nonumber
\end{cases}
\end{eqnarray*}

In his study, Ibănescu~\cite{IbanescuBC:phd} showed the presence of $\text{C}_{3}\text{H}_{5}\text{O}^{-}$ (57 amu), but we couldn't find it, possibly due to a very low cross section of the fragment production. In recent DEA studies with alcohols, authors have found the presence of $\text{O}^{-}$ ions, which are common products in the case of alcohols~\cite{Paul:2023,Chakraborty:2023,Ibanesu_Ethanol}. Interestingly, Ibănescu did not detect the presence of $\text{O}^{-}$ ions for propanols. In this study, we do find the presence of $\text{O}^{-}$ ions. The details of the detected products are discussed below.

\subsection{\label{subsec:h_production}Production of $\text{H}^{-}$ ions}

\begin{figure}[ht!]
\centering
\includegraphics[scale=0.57]{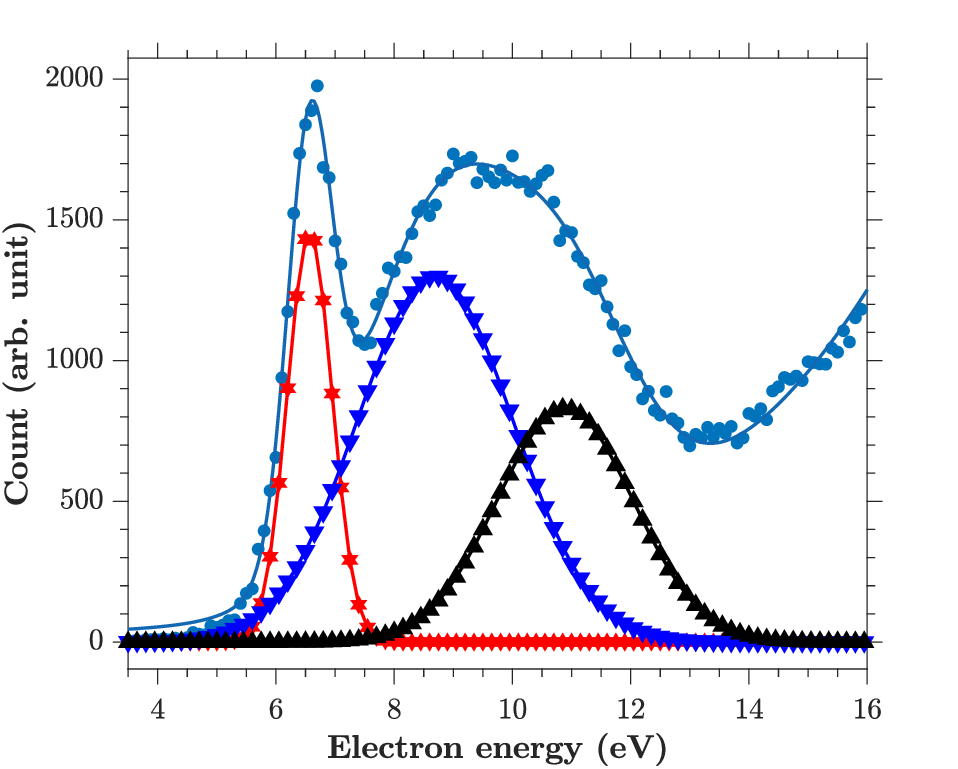}
\caption{\label{fig:H_partial_ion_yield}The ion yield of H$^-$ ions produced due to DEA to 1-propanol in the energy range of 3.5 to 16 eV. The blue scattered curve (indicated in \filleddot[matlabblue]{0.09}) denotes the experimental data, while the smooth blue line corresponds to the represents the cumulative fit of the ion yield (indicated in \textcolor{matlabblue}{\rule[0.5ex]{0.6cm}{1pt}}) with three Gaussian functions (illustrated in red, blue, and black). The fitting analysis revealed a sharp resonance at 6.5 eV (indicated in \textcolor{red}{\rule[0.5ex]{0.3cm}{1pt}}\filledStarOfDavid[red]{0.13}\textcolor{red}{\rule[0.5ex]{0.3cm}{1pt}}) and two broader resonances located at 8.7 eV (indicated in \textcolor{blue}{\rule[0.5ex]{0.3cm}{1pt}}\filleddowntriangle[blue]{0.13}\textcolor{blue}{\rule[0.5ex]{0.3cm}{1pt}}) and 10.9 eV (indicated in \textcolor{black}{\rule[0.5ex]{0.3cm}{1pt}}\filledtriangle[black]{0.13}\textcolor{black}{\rule[0.5ex]{0.3cm}{1pt}}). The raise in yield after 13.5 eV represents the ion-pair dissociation.}
\end{figure}

The ion yield of H$^-$ ions with respect to incident electron beam energy is presented in Fig.~\ref{fig:H_partial_ion_yield}. A broad range of overlapping resonances, spanning from 3.5 to 16 eV, is observed.  To locate the correct positions of these overlapping resonances, we opted to fit the H$^-$ ion yield using multiple Gaussian functions, as depicted in Fig.~\ref{fig:H_partial_ion_yield}. The fitted ion yield curve indicates the likely presence of three closely lying resonant bands in the 3.5 to 16 eV energy range, peaking at 6.5, 8.7 and 10.9 eV, respectively. However, because of the limited energy resolution (0.8 eV) of the electron beam and the finite width of the resonances, it is not feasible to separate them distinctly in the present study. Ibănescu \cite{IbanescuBC:phd} reported the ion-yield curve of $\text{H}^-$ ions, which is consistent in overall shape with our present results. Both spectra exhibit a sharp peak near 6.7 eV, a weaker peak around 8.0 eV, and a broad feature centered near 9.0 eV. The relative intensity of the first peak in Ibănescu’s data is slightly higher, a difference we cannot assess quantitatively since our spectrometer does not ensure total ion collection for $\text{H}^-$ ions. A comparable ion-yield profile was also observed in our earlier DEA studies on ethanol \cite{Paul:2023}.

The production of H$^{-}$ is possible via two-body dissociation channels. These dissociation channels are:

\begin{eqnarray}
\left(\text{CH}_3\text{CH}_2\text{CH}_2\text{OH}\right)^{-*}
\rightarrow &\text{H}^{-}+\text{CH}_{3}\text{CH}_{2}\text{CH}_{2}\text{O}\label{eqn:h_x1}\\
\rightarrow &\text{H}^{-}+\text{C}_{3}\text{H}_{6}\text{OH}\label{eqn:h_x2}
\end{eqnarray}

The channel~\ref{eqn:h_x1} suggest the dissociation happens via breaking the O$-$H bond and Ch.~\ref{eqn:h_x2} suggest the dissociation via breaking a C$-$H bond. The sharp rise in H$^{-}$ yield above 13.5~eV indicates the onset of dipolar dissociation in 1-propanol, where the incident electron transfers its energy to the molecule, promoting it to a superexcited state. Such states typically decay via dissociation, producing a pair of oppositely charged ions.

In related work, Ibănescu \textit{et al.} \cite{Ibanescu:2007} employed deuteration of the hydroxyl group to investigate the dissociation pathways. They attributed the resonances at 6.34 eV and 7.85 eV to O$-$H bond cleavage, while the resonance at 9.18 eV was assigned to hydrogen abstraction from the carbon chain. Due to the high resolution of the electron beam used in their study, the authors could distinctly observe the three resonances at 6.34, 7.85, and 9.18 eV for methanol. Comparable resonance patterns—typically consisting of one sharp feature and one broader structure—have been reported for a variety of alcohols, including ethanol, propanol, butanol, cyclopentanol, tetrahydrofuran-3-ol, (tetrahydrofuran-3-yl)methanol, (tetrahydrofuran-2-yl)methanol, and cis- and trans-cyclopentane-1,2-diols \cite{Ibanescu:2007,IbanescuBC:phd,Prabhudesai_2008_jcp,Prabhudesai:2005}. Across these systems, the $\text{H}^-$ yields exhibit similar structural characteristics, reinforcing the conclusion that the sharp resonance originates primarily from O$-$H bond cleavage, while the broad resonance results from a combination of O$-$H bond cleavage and hydrogen abstraction from the carbon chain. The data also indicate that variations in carbon chain length have little influence on the 6.4 eV resonance, but do affect the second resonance near 9.0 eV. This suggests that in DEA to 1-propanol, $\text{H}^-$ formation at 6.5~eV arises predominantly from O$-$H bond cleavage (Ch~\ref{eqn:h_x1}), whereas the 9.0~eV feature reflects O$-$H bond cleavage (Ch~\ref{eqn:h_x1}) with a smaller contribution from C$-$H bond cleavage (Ch~\ref{eqn:h_x2}). The authors assigned the first two resonances of 1-propanol, at 6.5~eV and 8.7~eV (Ch~\ref{eqn:h_x1}), to Feshbach resonances involving a hole in the out-of-plane ($n_{O}$) and in-plane ($\bar{n_{O}}$) oxygen lone-pair orbitals, respectively. The broad feature near 10.9~eV (Ch~\ref{eqn:h_x2}) is likely attributable to one or more Feshbach resonances arising from overlapping ionization bands of various C$-$H and C$-$C $\sigma$ orbitals lying above 12~eV.

The threshold calculations for the H$^-$ production channel are summarized in Table~\ref{tab:h_thresholds}. The results indicate that the cleavage of the O$-$H bond (3.33~eV) and the C$-$H bond (3.30~eV) occurs at nearly the same threshold energy. A possible three-body dissociation pathway, involving simultaneous cleavage of the O$-$H and C$-$H bonds, is also examined, with the corresponding threshold estimated to be slightly higher at 5.31~eV. The calculated results are in good agreement with the measurements. However, in the absence of information on the branching ratio, no definitive conclusion can be drawn regarding the dominant dissociation channel.

\begin{table}[!]
\caption{\label{tab:h_thresholds}Selected possible dissociation channels producing H$^-$ ions and calculated threshold values.}
\begin{ruledtabular}
\begin{tabular}{lcr}
Ch. no.                             & Dissociative products                 & Threshold energy (eV) \\
\colrule
(\ref{eqn:h_x1})                     & H$^-$ + CH$_3$CH$_2$CH$_2$O           & 3.33 \\
(\ref{eqn:h_x2})                     & H$^-$ + CH$_2$CH$_2$CH$_2$OH          & 3.30 \\
(\ref{eqn:h_x1}+\ref{eqn:h_x2})      & H$^-$ + CH$_2$CH$_2$CH$_2$O + H       & 5.31 \\
\end{tabular}
\end{ruledtabular}
\end{table}

\subsection{\label{subsec:oh_production}Production of OH$^{-}$ ions}

\begin{figure}[ht!]
\includegraphics[scale=0.57]{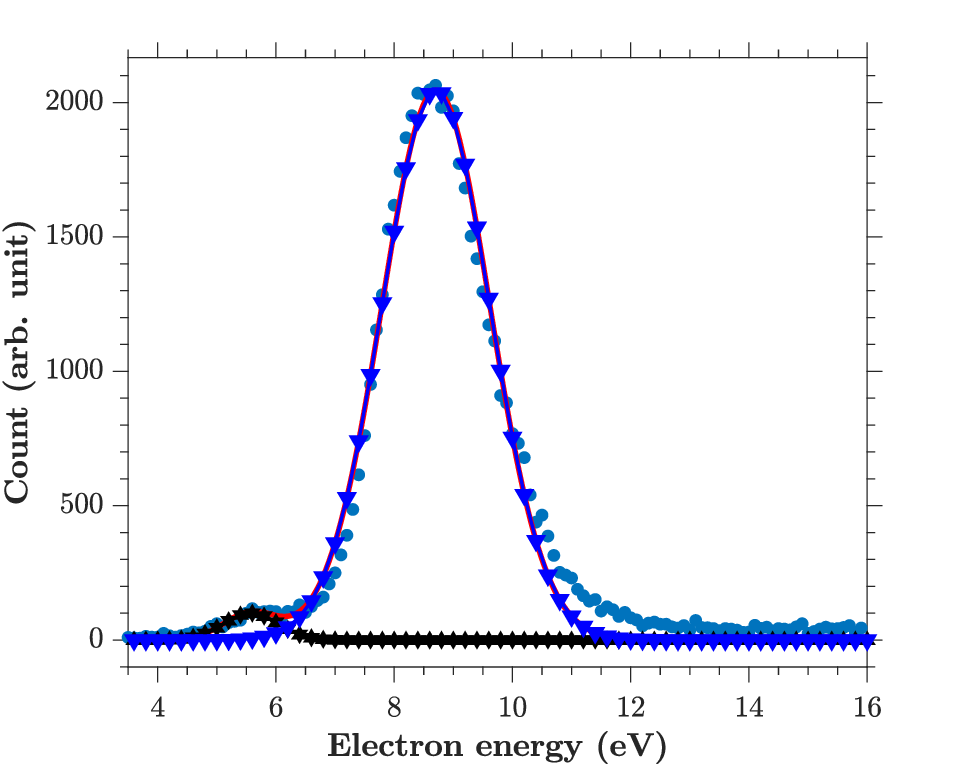}
\caption{\label{fig:OH_cs}The ion yield of OH$^{-}$ ions produced due to DEA to 1-propanol in 3.5 to 16 eV electron energy range. The experimental data is represented in scattered blue dots (indicated in \filleddot[matlabblue]{0.09}), the cumulative fit of the ion yield with two gaussian functions is represented in blue smooth line (indicated in \textcolor{red}{\rule[0.5ex]{0.6cm}{1pt}}). The fitting analysis reveals resonance features at 5.6 eV (indicated by \textcolor{black}{\rule[0.5ex]{0.3cm}{1pt}}\filledStarOfDavid[black]{0.13}\textcolor{black}{\rule[0.5ex]{0.3cm}{1pt}}) and 8.7 eV (indicated by \textcolor{blue}{\rule[0.5ex]{0.3cm}{1pt}}\filleddowntriangle[blue]{0.13}\textcolor{blue}{\rule[0.5ex]{0.3cm}{1pt}}).}
\end{figure}

Fig.~\ref{fig:OH_cs} illustrates the ion-yield curve of OH$^-$ ions resulting from DEA to 1-propanol. A broad resonance is observed around 8.7 eV, accompanied by a smaller hump near 5.6 eV. Due to the broad nature of the 8.7 eV resonance, the contribution of other, lower cross-section resonances in this energy region cannot be ruled out. The 8.7 eV resonance is likely attributed to one or more Feshbach resonances, which arise due to the overlapping features observed in the photoelectron spectrum, resulting from ionization of multiple C$-$H and C$-$C $\sigma$ orbitals above 12 eV \cite{IbanescuBC:phd}. The formation of OH$^-$ ions during the DEA process occurs via the dissociation of the C$_3$H$_7$$-$OH. The possible dissociation channels can be represented as follows:

\begin{eqnarray}
\left(\text{CH}_3\text{CH}_2\text{CH}_2\text{OH}\right)^{-*} \rightarrow &\text{OH}^-+\text{CH}_3\text{CH}_2\text{CH}_2 \label{eqn:oh_x1} \\ 
\rightarrow &\text{OH}^-+\text{C}_3\text{H}_6 + \text{H} \label{eqn:oh_x2}
\end{eqnarray}

Ch.~\ref{eqn:oh_x1} is a simple two-body dissociation that breaks the C$-$O bond, and the excess electron is captured by the OH moiety. This dissociation channel can be compared with findings reported in the literature. Ibanescu \emph{et al.} conducted a series of DEA experiments on various alcohols and ethers over the years \cite{Ibanescu:2007,IbanescuBC:phd}. Their detailed investigations focused on different DEA resonances responsible for the dissociation of specific bonds, such as C$-$H, O$-$H, C$-$O, and C$-$C. Interestingly, their findings indicate that the energy of the Feshbach resonance involved in the DEA process does not depend on the nature of the parent molecule or the specific negative ions formed during dissociation. Rather, it is influenced by the type of neutral conjugate produced in the process. Such selectivity within a dense manifold of highly excited TNIs is unexpected, as individual states are undoubtedly strongly vibronically coupled, allowing transitions between them via numerous conical intersections. Moreover, this selectivity has been observed for a variety of radicals, including methyl, ethyl, propyl, and n-butyl species. These observations suggest a consistent connection between the energy of the Feshbach resonance involved in DEA and the resulting neutral conjugates, regardless of the parent molecule. To explore this connection, Chakraborty \emph{et al.} \cite{Chakraborty:2023} investigated the ethanol molecule—the simplest candidate in this group that yields the ethyl radical—which exhibits a resonance near 9.5 eV. They concluded that a $p^2$-Feshbach resonance is involved in the dissociation process. In the present study, the OH$^-$ ion is accompanied by the propyl radical, and the observed resonance near 8.7 eV aligns well with previous findings \cite{IbanescuBC:phd}.

In Ch.~\ref{eqn:oh_x2}, an additional C$-$H bond breaks along with the C$-$O bond, resulting in the formation of an H$^\cdot$ radical and C$_3$H$_6$. Since C$_3$H$_6$ is a stable compound, this makes Ch.~\ref{eqn:oh_x2} a plausible dissociation pathway in this region, provided that the internal energy of C$_3$H$_7$ is sufficiently high. Therefore, Ch.~\ref{eqn:oh_x2} could proceed either via sequential dissociation, with C$_3$H$_7$ from Ch.~\ref{eqn:oh_x1} acting as an intermediate, or through a concerted three-body dissociation involving simultaneous cleavage of the C$-$O and C$-$H bonds. In a previous study on ethanol, a molecule from a similar group, comparable dissociation channels were observed in which H$^\cdot$ radicals were reported during OH$^-$ formation \cite{Paul:2023}. Based on the internal energy of the propyl radical, three-body or higher-order dissociation may be possible. However, in the present study, we cannot comment on this, as the kinetic energies of the fragments are unknown.

To gain deeper insight into the dissociation pathways, threshold energy calculations are performed for the OH$^-$ dissociation channel, and the results are summarized in Table~\ref{tab:oh_thresholds}. Cleavage of only the C$-$O bond requires substantially less energy than the simultaneous breaking of the C$-$O and C$-$H bonds, which leads to the formation of a stable CH$_2$CH$_2$CH$_2$ and H$^\cdot$ radicals. Since both thresholds lie well below the experimentally observed peak, the possibility that multiple channels contribute to the observed features cannot be excluded.

\begin{table}[!]
\caption{\label{tab:oh_thresholds}Selected possible dissociation channels producing OH$^-$ ions and calculated threshold values.}
\begin{ruledtabular}
\begin{tabular}{lcr}
Ch. no.             & Dissociative products                 & Threshold energy (eV) \\
\colrule
(\ref{eqn:oh_x1})   & OH$^-$ + CH$_3$CH$_2$CH$_2$           & 1.87 \\
(\ref{eqn:oh_x2})   & OH$^-$ + CH$_2$CH$_2$CH$_2$ + H       & 3.74 \\
\end{tabular}
\end{ruledtabular}
\end{table}

\subsection{\label{subsec:c3h7o_production}H-loss channel (C$_3$H$_7$O$^-$ ion production)}

\begin{figure}[ht!]
\centering
\includegraphics[scale=0.57]{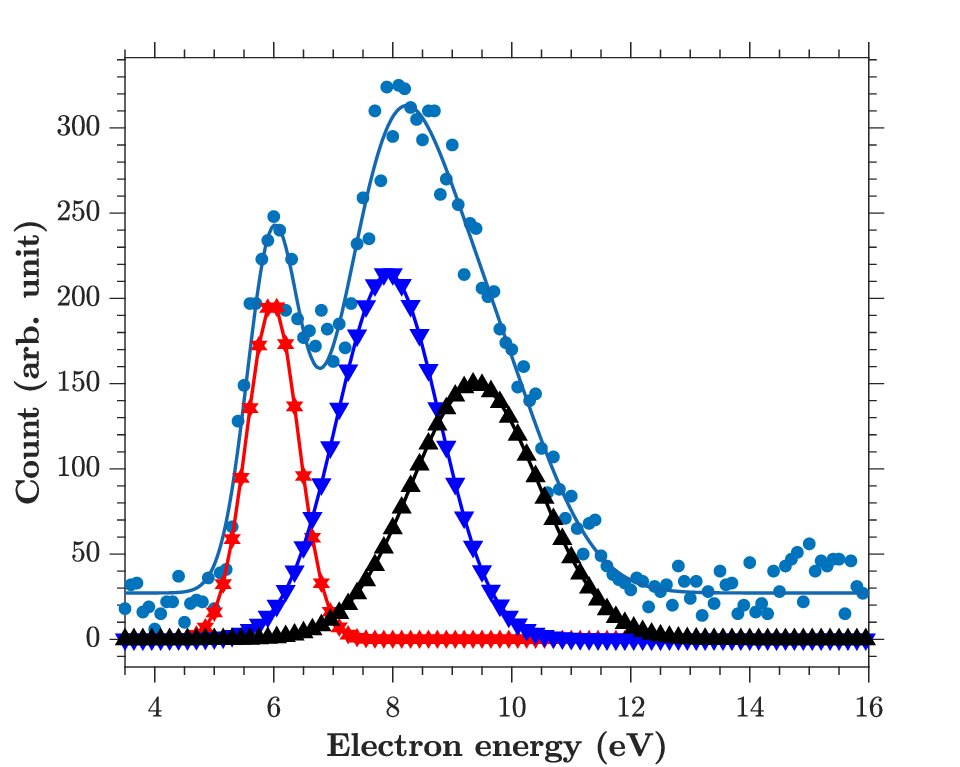}
\caption{\label{fig:C3H7O_cs}The ion yield of C$_3$H$_7$O$^-$ ions produced due to DEA to 1-propanol in 3.5 to 16 eV electron energy range, where the scattered blue dots (marked by \filleddot[matlabblue]{0.09}) represent the experimental data and the smooth blue line corresponds to the cumulative fit of the ion yield (indicated in \textcolor{matlabblue}{\rule[0.5ex]{0.6cm}{1pt}}) with three gaussian functions (illustrated in red, blue and black). A background contribution is present in the spectrum. The fitting analysis revealed resonance features at approximately 6.0 eV (marked by \textcolor{red}{\rule[0.5ex]{0.3cm}{1pt}}\filledStarOfDavid[red]{0.13}\textcolor{red}{\rule[0.5ex]{0.3cm}{1pt}}), 8.0 eV (marked by \textcolor{blue}{\rule[0.5ex]{0.3cm}{1pt}}\filleddowntriangle[blue]{0.13}\textcolor{blue}{\rule[0.5ex]{0.3cm}{1pt}}), and 9.4 eV (marked by \textcolor{black}{\rule[0.5ex]{0.3cm}{1pt}}\filledtriangle[black]{0.13}\textcolor{black}{\rule[0.5ex]{0.3cm}{1pt}}).}
\end{figure}

The yield of C$_3$H$_7$O$^-$ ions as a function of incident electron energy from DEA to 1-propanol is depicted in Fig.~\ref{fig:C3H7O_cs}. The resonance observed around 6.0 eV, along with a broad resonance extending from 7 to 11 eV, with a maximum near 8.1 eV. A similar ion yield profile was reported in a previous study \cite{IbanescuBC:phd}. These resonant features indicate the presence of Feshbach resonances, consistent with those identified for other fragment ions. The formation of the C$_3$H$_7$O$^-$ ion occurs via a two-body dissociation pathway, either through the cleavage of a C$-$H bond from the propyl group or via O$-$H bond dissociation at the hydroxyl site. The two possible dissociation channels can be represented as follows:

\begin{eqnarray}
\left(\text{CH}_3\text{CH}_2\text{CH}_2\text{OH}\right)^{-*} \rightarrow &\text{C}_3\text{H}_6\text{OH}^- + \text{H} \label{eqn:c3h7o_x1}\\ 
\rightarrow &\text{C}_3\text{H}_7\text{O}^- + \text{H} \label{eqn:c3h7o_x2}
\end{eqnarray}

Previous studies on structurally similar alcohols suggest that H-loss from the $-$OH group, i.e., Ch.~\ref{eqn:c3h7o_x2}, is the more favorable pathway \cite{Ibanescu:2007,Paul:2023}. This process is essentially the conjugate dissociation of H$^-$ formation (Ch.~\ref{eqn:h_x1}), in which resonances were observed near 6.5, and 9.5 eV. The similarity in the resonance energies and the overall shape of the ion yield curves suggests that both the H$^-$ (Ch.~\ref{eqn:h_x1}) and C$_3$H$_7$O$^-$ (Ch.~\ref{eqn:c3h7o_x2}) ions originate from O$-$H bond cleavage, which appears to be the dominant dissociation pathway in this case. However, a definitive conclusion requires further study using a deuterated analogue of the molecule.

The resonance near 6 eV can be assigned to a Feshbach resonance involving a hole in the out-of-plane ($n_O$) lone pair orbital of the oxygen atom \cite{IbanescuBC:phd}. The broad feature centered around 8.1 eV is most likely associated with one or several Feshbach resonances arising from overlapping bands in the photoelectron spectrum, corresponding to ionization from various C$-$H and C$-$C $\sigma$ orbitals above 12 eV, as discussed in the previous sections. The threshold energy calculations for the dissociation pathway involving O$-$H and C$-$H bond cleavage are consistent with the experimental observation. The calculated threshold energies are summarized in Table~\ref{tab:c3h7o_thresholds}.

\begin{table}[!]
\caption{\label{tab:c3h7o_thresholds}Selected possible dissociation channels producing C$_3$H$_7$O$^-$ ions and calculated threshold values.}
\begin{ruledtabular}
\begin{tabular}{lcr}
Ch. no.                 & Dissociative products                 & Threshold energy (eV) \\
\colrule
(\ref{eqn:c3h7o_x1})    & CH$_2$CH$_2$CH$_2$OH$^-$ + H          & 3.58 \\
(\ref{eqn:c3h7o_x2})    & CH$_3$CH$_2$CH$_2$O$^-$ + H           & 2.54 \\
\end{tabular}
\end{ruledtabular}
\end{table}

\subsection{\label{subsec:o_production}Production of $\text{O}^{-}$ ions}

\begin{figure}[ht!]
\centering
\includegraphics[scale=0.57]{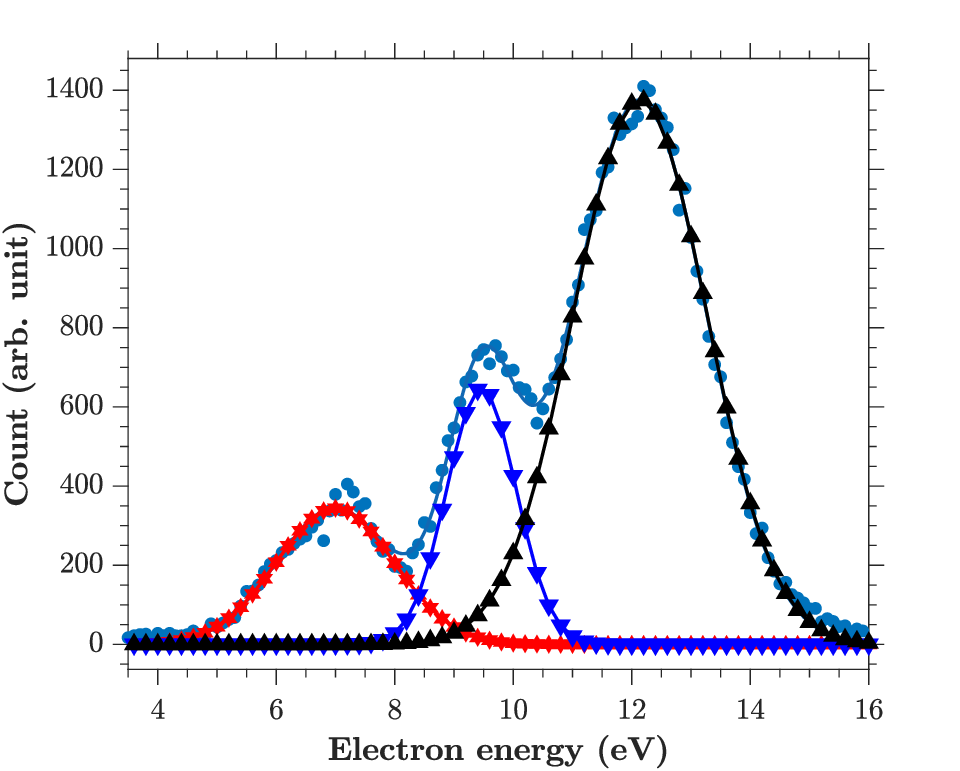}
\caption{\label{fig:O_cs}The ion yield of O$^-$ ions resulting from DEA to 1-propanol is measured in the 3.5--16~eV electron energy range. The scattered blue curve (indicated by \filleddot[matlabblue]{0.09}) represents the experimental data, while smooth blue line corresponds to the fitted curve (indicated in \textcolor{matlabblue}{\rule[0.5ex]{0.6cm}{1pt}}). The fitting analysis revealed resonance features at approximately 6.9 eV (marked by \textcolor{red}{\rule[0.5ex]{0.3cm}{1pt}}\filledStarOfDavid[red]{0.13}\textcolor{red}{\rule[0.5ex]{0.3cm}{1pt}}), 9.5 eV (marked by \textcolor{blue}{\rule[0.5ex]{0.3cm}{1pt}}\filleddowntriangle[blue]{0.13}\textcolor{blue}{\rule[0.5ex]{0.3cm}{1pt}}), and 12.1 eV (marked by \textcolor{black}{\rule[0.5ex]{0.3cm}{1pt}}\filledtriangle[black]{0.13}\textcolor{black}{\rule[0.5ex]{0.3cm}{1pt}}). This yield curve closely resembles the O$^-$/H$_2$O ion yield reported in Refs.~\cite{Fedor_2006,Rawat_2007}.} 
\end{figure}

The yield of O$^-$ ions as a function of the incident electron energy resulting from DEA to 1-propanol is shown in Fig.~\ref{fig:O_cs}. Three distinct resonances are observed around 6.9, 9.5, and 12.1 eV. It is worth noting that the measured O$^-$ ion yield profile is closely resembles the O$^-$ ion yield reported for water \cite{Fedor_2006,Rawat_2007}. This observation suggests that the signal could, at least in part, originate from residual water impurities that are difficult to exclude entirely from the liquid sample. Nevertheless, it is also important to consider that the appearance of O$^-$ ions is not unexpected in alcohols. Previous studies on ethanol, a structurally related molecule, have shown distinct O$^-$ resonances around 6.7, 9.8, and 11.7 eV \cite{Paul:2023}. Such behavior has been rationalized in terms of functional group–dependent dissociative electron attachment (DEA), a concept further supported by earlier investigations on different alcohols \cite{Prabhudesai:2005,Prabhudesai_2008_jcp}. Based on this argument, it is reasonable to assume that in 1-propanol, similar resonance structures should occur in the vicinity of 7, 9.5, and 12.1 eV, even though the distinctive features are not clearly resolved in the present measurements due to masking effects from water contamination. Despite repeated experimental attempts, complete removal of water impurities from the target sample was not feasible, and therefore the O$^-$ ion yield reported here likely represents a convolution of contributions from both 1-propanol and trace water.

\section{\label{sec4:conclusion}Conclusion}

This study presents a comprehensive investigation of the DEA dynamics of 1-propanol over an incident electron energy range of 3.5–16 eV, revealing four fragment anions— H$^-$, O$^-$, OH$^-$, and C$_3$H$_7$O$^-$—and their associated resonance structures. Differential cross sections for H$^-$, OH$^-$, and C$_3$H$_7$O$^-$ are measured, and the observed dissociation pathways are discussed. The resonance features, largely assigned to Feshbach resonances, elucidate key bond-cleavage mechanisms, with O$-$H bond dissociation emerging as a dominant channel for both H$^-$ and C$_3$H$_7$O$^-$ production. Our experimental findings, combined with theoretical calculations, provide insight into the different dissociation channels. These findings have practical implications for optimizing alcohol-based fuels in combustion and plasma-assisted ignition systems, as well as for predicting alcohol behavior in ionizing environments.

\begin{acknowledgements}
S. Ghosh deeply appreciates the ``Indian Institute of Science Education and Research (IISER) Kolkata'' for providing financial assistance and research facilities. A. Paul acknowledges the ``Council of Scientific and Industrial Research (CSIR)'' for Ph.D. fellowship.
\end{acknowledgements}

\bibliography{main}

\providecommand{\noopsort}[1]{}\providecommand{\singleletter}[1]{#1}%
\begin{thebibliography}{50}%
\makeatletter
\providecommand \@ifxundefined [1]{%
 \@ifx{#1\undefined}
}%
\providecommand \@ifnum [1]{%
 \ifnum #1\expandafter \@firstoftwo
 \else \expandafter \@secondoftwo
 \fi
}%
\providecommand \@ifx [1]{%
 \ifx #1\expandafter \@firstoftwo
 \else \expandafter \@secondoftwo
 \fi
}%
\providecommand \natexlab [1]{#1}%
\providecommand \enquote  [1]{``#1''}%
\providecommand \bibnamefont  [1]{#1}%
\providecommand \bibfnamefont [1]{#1}%
\providecommand \citenamefont [1]{#1}%
\providecommand \href@noop [0]{\@secondoftwo}%
\providecommand \href [0]{\begingroup \@sanitize@url \@href}%
\providecommand \@href[1]{\@@startlink{#1}\@@href}%
\providecommand \@@href[1]{\endgroup#1\@@endlink}%
\providecommand \@sanitize@url [0]{\catcode `\\12\catcode `\$12\catcode
  `\&12\catcode `\#12\catcode `\^12\catcode `\_12\catcode `\%12\relax}%
\providecommand \@@startlink[1]{}%
\providecommand \@@endlink[0]{}%
\providecommand \url  [0]{\begingroup\@sanitize@url \@url }%
\providecommand \@url [1]{\endgroup\@href {#1}{\urlprefix }}%
\providecommand \urlprefix  [0]{URL }%
\providecommand \Eprint [0]{\href }%
\providecommand \doibase [0]{https://doi.org/}%
\providecommand \selectlanguage [0]{\@gobble}%
\providecommand \bibinfo  [0]{\@secondoftwo}%
\providecommand \bibfield  [0]{\@secondoftwo}%
\providecommand \translation [1]{[#1]}%
\providecommand \BibitemOpen [0]{}%
\providecommand \bibitemStop [0]{}%
\providecommand \bibitemNoStop [0]{.\EOS\space}%
\providecommand \EOS [0]{\spacefactor3000\relax}%
\providecommand \BibitemShut  [1]{\csname bibitem#1\endcsname}%
\let\auto@bib@innerbib\@empty
\bibitem [{\citenamefont {Fabrikant}\ \emph {et~al.}(2017)\citenamefont
  {Fabrikant}, \citenamefont {Eden}, \citenamefont {Mason},\ and\ \citenamefont
  {Fedor}}]{Fabrikant_Review}%
  \BibitemOpen
  \bibfield  {author} {\bibinfo {author} {\bibfnamefont {I.~I.}\ \bibnamefont
  {Fabrikant}}, \bibinfo {author} {\bibfnamefont {S.}~\bibnamefont {Eden}},
  \bibinfo {author} {\bibfnamefont {N.~J.}\ \bibnamefont {Mason}},\ and\
  \bibinfo {author} {\bibfnamefont {J.}~\bibnamefont {Fedor}},\ }\bibfield
  {title} {\bibinfo {title} {{Chapter Nine - Recent Progress in Dissociative
  Electron Attachment: From Diatomics to Biomolecules}}\ }(\bibinfo
  {publisher} {Academic Press},\ \bibinfo {year} {2017})\ pp.\ \bibinfo {pages}
  {545--657}\BibitemShut {NoStop}%
\bibitem [{\citenamefont {Ingólfsson}(2019)}]{ingolfsson_low-energy_2019}%
  \BibitemOpen
  \bibinfo {editor} {\bibfnamefont {O.}~\bibnamefont {Ingólfsson}},\ ed.,\
  \href {https://doi.org/10.1201/9780429058820} {\emph {\bibinfo {title}
  {Low-Energy Electrons: Fundamentals and Applications}}}\ (\bibinfo
  {publisher} {Jenny Stanford Publishing},\ \bibinfo {address} {New York},\
  \bibinfo {year} {2019})\BibitemShut {NoStop}%
\bibitem [{\citenamefont {Illenberger}\ and\ \citenamefont
  {Momigny}(1992)}]{Illenberger_book}%
  \BibitemOpen
  \bibfield  {author} {\bibinfo {author} {\bibfnamefont {E.}~\bibnamefont
  {Illenberger}}\ and\ \bibinfo {author} {\bibfnamefont {J.}~\bibnamefont
  {Momigny}},\ }\href {https://doi.org/10.1007/978-3-662-07383-4} {\emph
  {\bibinfo {title} {{Gaseous Molecular Ions: An Introduction to Elementary
  Processes Induced by Ionization}}}},\ Topics in Physical Chemistry\ (\bibinfo
   {publisher} {Steinkopff Heidelberg},\ \bibinfo {year} {1992})\BibitemShut
  {NoStop}%
\bibitem [{\citenamefont {Schulz}(1973)}]{Schulz}%
  \BibitemOpen
  \bibfield  {author} {\bibinfo {author} {\bibfnamefont {G.~J.}\ \bibnamefont
  {Schulz}},\ }\bibfield  {title} {\bibinfo {title} {{Resonances in Electron
  Impact on Diatomic Molecules}},\ }\href
  {https://doi.org/10.1103/RevModPhys.45.423} {\bibfield  {journal} {\bibinfo
  {journal} {Rev. Mod. Phys.}\ }\textbf {\bibinfo {volume} {45}},\ \bibinfo
  {pages} {423} (\bibinfo {year} {1973})}\BibitemShut {NoStop}%
\bibitem [{\citenamefont {Chakraborty}\ \emph {et~al.}(2019)\citenamefont
  {Chakraborty}, \citenamefont {Giri},\ and\ \citenamefont
  {Nandi}}]{Chakraborty_Ammonia}%
  \BibitemOpen
  \bibfield  {author} {\bibinfo {author} {\bibfnamefont {D.}~\bibnamefont
  {Chakraborty}}, \bibinfo {author} {\bibfnamefont {A.}~\bibnamefont {Giri}},\
  and\ \bibinfo {author} {\bibfnamefont {D.}~\bibnamefont {Nandi}},\ }\bibfield
   {title} {\bibinfo {title} {{Dissociation dynamics in low energy electron
  attachment to ammonia using velocity slice imaging}},\ }\href
  {https://doi.org/10.1039/C9CP03973B} {\bibfield  {journal} {\bibinfo
  {journal} {Phys. Chem. Chem. Phys.}\ }\textbf {\bibinfo {volume} {21}},\
  \bibinfo {pages} {21908} (\bibinfo {year} {2019})}\BibitemShut {NoStop}%
\bibitem [{\citenamefont {Adaniya}\ \emph {et~al.}(2009)\citenamefont
  {Adaniya}, \citenamefont {Rudek}, \citenamefont {Osipov}, \citenamefont
  {Haxton}, \citenamefont {Weber}, \citenamefont {Rescigno}, \citenamefont
  {McCurdy},\ and\ \citenamefont {Belkacem}}]{Adaniya_Water}%
  \BibitemOpen
  \bibfield  {author} {\bibinfo {author} {\bibfnamefont {H.}~\bibnamefont
  {Adaniya}}, \bibinfo {author} {\bibfnamefont {B.}~\bibnamefont {Rudek}},
  \bibinfo {author} {\bibfnamefont {T.}~\bibnamefont {Osipov}}, \bibinfo
  {author} {\bibfnamefont {D.~J.}\ \bibnamefont {Haxton}}, \bibinfo {author}
  {\bibfnamefont {T.}~\bibnamefont {Weber}}, \bibinfo {author} {\bibfnamefont
  {T.~N.}\ \bibnamefont {Rescigno}}, \bibinfo {author} {\bibfnamefont {C.~W.}\
  \bibnamefont {McCurdy}},\ and\ \bibinfo {author} {\bibfnamefont
  {A.}~\bibnamefont {Belkacem}},\ }\bibfield  {title} {\bibinfo {title}
  {{Imaging the Molecular Dynamics of Dissociative Electron Attachment to
  Water}},\ }\href {https://doi.org/10.1103/PhysRevLett.103.233201} {\bibfield
  {journal} {\bibinfo  {journal} {Phys. Rev. Lett.}\ }\textbf {\bibinfo
  {volume} {103}},\ \bibinfo {pages} {233201} (\bibinfo {year}
  {2009})}\BibitemShut {NoStop}%
\bibitem [{\citenamefont {Chakraborty}\ \emph {et~al.}(2023)\citenamefont
  {Chakraborty}, \citenamefont {Slaughter},\ and\ \citenamefont
  {Ptasinska}}]{Chakraborty:2023}%
  \BibitemOpen
  \bibfield  {author} {\bibinfo {author} {\bibfnamefont {D.}~\bibnamefont
  {Chakraborty}}, \bibinfo {author} {\bibfnamefont {D.~S.}\ \bibnamefont
  {Slaughter}},\ and\ \bibinfo {author} {\bibfnamefont {S.}~\bibnamefont
  {Ptasinska}},\ }\bibfield  {title} {\bibinfo {title} {{Dynamics of resonant
  low-energy electron attachment to ethanol-producing hydroxide anions}},\
  }\href {https://doi.org/10.1103/PhysRevA.108.052806} {\bibfield  {journal}
  {\bibinfo  {journal} {Phys. Rev. A}\ }\textbf {\bibinfo {volume} {108}},\
  \bibinfo {pages} {052806} (\bibinfo {year} {2023})}\BibitemShut {NoStop}%
\bibitem [{\citenamefont {Prabhudesai}\ \emph {et~al.}(2005)\citenamefont
  {Prabhudesai}, \citenamefont {Kelkar}, \citenamefont {Nandi},\ and\
  \citenamefont {Krishnakumar}}]{Prabhudesai:2005}%
  \BibitemOpen
  \bibfield  {author} {\bibinfo {author} {\bibfnamefont {V.~S.}\ \bibnamefont
  {Prabhudesai}}, \bibinfo {author} {\bibfnamefont {A.~H.}\ \bibnamefont
  {Kelkar}}, \bibinfo {author} {\bibfnamefont {D.}~\bibnamefont {Nandi}},\ and\
  \bibinfo {author} {\bibfnamefont {E.}~\bibnamefont {Krishnakumar}},\
  }\bibfield  {title} {\bibinfo {title} {{Functional Group Dependent Site
  Specific Fragmentation of Molecules by Low Energy Electrons}},\ }\href
  {https://doi.org/10.1103/PhysRevLett.95.143202} {\bibfield  {journal}
  {\bibinfo  {journal} {Phys. Rev. Lett.}\ }\textbf {\bibinfo {volume} {95}},\
  \bibinfo {pages} {143202} (\bibinfo {year} {2005})}\BibitemShut {NoStop}%
\bibitem [{\citenamefont {Ibănescu}\ \emph {et~al.}(2007)\citenamefont
  {Ibănescu}, \citenamefont {May}, \citenamefont {Monney},\ and\ \citenamefont
  {Allan}}]{Ibanescu:2007}%
  \BibitemOpen
  \bibfield  {author} {\bibinfo {author} {\bibfnamefont {B.~C.}\ \bibnamefont
  {Ibănescu}}, \bibinfo {author} {\bibfnamefont {O.}~\bibnamefont {May}},
  \bibinfo {author} {\bibfnamefont {A.}~\bibnamefont {Monney}},\ and\ \bibinfo
  {author} {\bibfnamefont {M.}~\bibnamefont {Allan}},\ }\bibfield  {title}
  {\bibinfo {title} {{Electron-induced chemistry of alcohols}},\ }\href
  {https://doi.org/10.1039/B704656A} {\bibfield  {journal} {\bibinfo  {journal}
  {Phys. Chem. Chem. Phys.}\ }\textbf {\bibinfo {volume} {9}},\ \bibinfo
  {pages} {3163} (\bibinfo {year} {2007})}\BibitemShut {NoStop}%
\bibitem [{\citenamefont {Paul}\ \emph {et~al.}(2023)\citenamefont {Paul},
  \citenamefont {Ghosh},\ and\ \citenamefont {Nandi}}]{Paul:2023}%
  \BibitemOpen
  \bibfield  {author} {\bibinfo {author} {\bibfnamefont {A.}~\bibnamefont
  {Paul}}, \bibinfo {author} {\bibfnamefont {S.}~\bibnamefont {Ghosh}},\ and\
  \bibinfo {author} {\bibfnamefont {D.}~\bibnamefont {Nandi}},\ }\bibfield
  {title} {\bibinfo {title} {{Fragmentation dynamics and absolute dissociative
  electron attachment cross sections in the low energy electron collision with
  ethanol}},\ }\href {https://doi.org/10.1039/D3CP03601D} {\bibfield  {journal}
  {\bibinfo  {journal} {Phys. Chem. Chem. Phys.}\ }\textbf {\bibinfo {volume}
  {25}},\ \bibinfo {pages} {28263} (\bibinfo {year} {2023})}\BibitemShut
  {NoStop}%
\bibitem [{\citenamefont {Nixon}\ \emph {et~al.}(2016)\citenamefont {Nixon},
  \citenamefont {Pires}, \citenamefont {Neves}, \citenamefont {Duque},
  \citenamefont {Jones}, \citenamefont {Brunger},\ and\ \citenamefont
  {Lopes}}]{NIXON:2016}%
  \BibitemOpen
  \bibfield  {author} {\bibinfo {author} {\bibfnamefont {K.}~\bibnamefont
  {Nixon}}, \bibinfo {author} {\bibfnamefont {W.}~\bibnamefont {Pires}},
  \bibinfo {author} {\bibfnamefont {R.}~\bibnamefont {Neves}}, \bibinfo
  {author} {\bibfnamefont {H.}~\bibnamefont {Duque}}, \bibinfo {author}
  {\bibfnamefont {D.}~\bibnamefont {Jones}}, \bibinfo {author} {\bibfnamefont
  {M.}~\bibnamefont {Brunger}},\ and\ \bibinfo {author} {\bibfnamefont
  {M.}~\bibnamefont {Lopes}},\ }\bibfield  {title} {\bibinfo {title} {{Electron
  impact ionisation and fragmentation of methanol and ethanol}},\ }\href
  {https://doi.org/https://doi.org/10.1016/j.ijms.2016.05.006} {\bibfield
  {journal} {\bibinfo  {journal} {Int. J. Mass Spectrom.}\ }\textbf {\bibinfo
  {volume} {404}},\ \bibinfo {pages} {48} (\bibinfo {year} {2016})}\BibitemShut
  {NoStop}%
\bibitem [{\citenamefont {Szmytkowski}\ and\ \citenamefont
  {Krzysztofowicz}(1995)}]{C_Szmytkowski_1995}%
  \BibitemOpen
  \bibfield  {author} {\bibinfo {author} {\bibfnamefont {C.}~\bibnamefont
  {Szmytkowski}}\ and\ \bibinfo {author} {\bibfnamefont {A.~M.}\ \bibnamefont
  {Krzysztofowicz}},\ }\bibfield  {title} {\bibinfo {title} {{Electron
  scattering from isoelectronic, Ne=18, CH$_3$X molecules (X=F, OH, NH$_2$ and
  CH$_3$)}},\ }\href {https://doi.org/10.1088/0953-4075/28/19/014} {\bibfield
  {journal} {\bibinfo  {journal} {J. Phys. B: At. Mol. Opt. Phys.}\ }\textbf
  {\bibinfo {volume} {28}},\ \bibinfo {pages} {4291} (\bibinfo {year}
  {1995})}\BibitemShut {NoStop}%
\bibitem [{\citenamefont {Vinodkumar}\ \emph {et~al.}(2008)\citenamefont
  {Vinodkumar}, \citenamefont {Limbachiya}, \citenamefont {Joshipura},
  \citenamefont {Vaishnav},\ and\ \citenamefont
  {Gangopadhyay}}]{M_Vinodkumar_2008}%
  \BibitemOpen
  \bibfield  {author} {\bibinfo {author} {\bibfnamefont {M.}~\bibnamefont
  {Vinodkumar}}, \bibinfo {author} {\bibfnamefont {C.}~\bibnamefont
  {Limbachiya}}, \bibinfo {author} {\bibfnamefont {K.~N.}\ \bibnamefont
  {Joshipura}}, \bibinfo {author} {\bibfnamefont {B.}~\bibnamefont
  {Vaishnav}},\ and\ \bibinfo {author} {\bibfnamefont {S.}~\bibnamefont
  {Gangopadhyay}},\ }\bibfield  {title} {\bibinfo {title} {{Computation of
  total electron scattering cross sections for molecules of astrophysical
  relevance}},\ }\href {https://doi.org/10.1088/1742-6596/115/1/012013}
  {\bibfield  {journal} {\bibinfo  {journal} {J. Phys.: Conf. Ser.}\ }\textbf
  {\bibinfo {volume} {115}},\ \bibinfo {pages} {012013} (\bibinfo {year}
  {2008})}\BibitemShut {NoStop}%
\bibitem [{\citenamefont {Silva}\ \emph {et~al.}(2009)\citenamefont {Silva},
  \citenamefont {Tejo}, \citenamefont {Muse}, \citenamefont {Romero},
  \citenamefont {Khakoo},\ and\ \citenamefont {Lopes}}]{Silva_2010}%
  \BibitemOpen
  \bibfield  {author} {\bibinfo {author} {\bibfnamefont {D.~G.~M.}\
  \bibnamefont {Silva}}, \bibinfo {author} {\bibfnamefont {T.}~\bibnamefont
  {Tejo}}, \bibinfo {author} {\bibfnamefont {J.}~\bibnamefont {Muse}}, \bibinfo
  {author} {\bibfnamefont {D.}~\bibnamefont {Romero}}, \bibinfo {author}
  {\bibfnamefont {M.~A.}\ \bibnamefont {Khakoo}},\ and\ \bibinfo {author}
  {\bibfnamefont {M.~C.~A.}\ \bibnamefont {Lopes}},\ }\bibfield  {title}
  {\bibinfo {title} {{Total electron scattering cross sections for methanol and
  ethanol at intermediate energies}},\ }\href
  {https://doi.org/10.1088/0953-4075/43/1/015201} {\bibfield  {journal}
  {\bibinfo  {journal} {J. Phys. B: At. Mol. Opt. Phys.}\ }\textbf {\bibinfo
  {volume} {43}},\ \bibinfo {pages} {015201} (\bibinfo {year}
  {2009})}\BibitemShut {NoStop}%
\bibitem [{\citenamefont {Tan}\ and\ \citenamefont {Wang}(2011)}]{TAN:2011}%
  \BibitemOpen
  \bibfield  {author} {\bibinfo {author} {\bibfnamefont {X.-M.}\ \bibnamefont
  {Tan}}\ and\ \bibinfo {author} {\bibfnamefont {D.-H.}\ \bibnamefont {Wang}},\
  }\bibfield  {title} {\bibinfo {title} {{Total cross sections for electron
  scattering from CH$_3$OH and CH$_3$CH$_2$OH molecules in the energy range
  from 10 to 1000 eV}},\ }\href
  {https://doi.org/https://doi.org/10.1016/j.nimb.2011.03.008} {\bibfield
  {journal} {\bibinfo  {journal} {NIM-B}\ }\textbf {\bibinfo {volume} {269}},\
  \bibinfo {pages} {1094} (\bibinfo {year} {2011})}\BibitemShut {NoStop}%
\bibitem [{\citenamefont {Vinodkumar}\ \emph {et~al.}(2013)\citenamefont
  {Vinodkumar}, \citenamefont {Limbachiya}, \citenamefont {Barot},\ and\
  \citenamefont {Mason}}]{Vinodkumar:2013}%
  \BibitemOpen
  \bibfield  {author} {\bibinfo {author} {\bibfnamefont {M.}~\bibnamefont
  {Vinodkumar}}, \bibinfo {author} {\bibfnamefont {C.}~\bibnamefont
  {Limbachiya}}, \bibinfo {author} {\bibfnamefont {A.}~\bibnamefont {Barot}},\
  and\ \bibinfo {author} {\bibfnamefont {N.}~\bibnamefont {Mason}},\ }\bibfield
   {title} {\bibinfo {title} {{Computation of electron-impact rotationally
  elastic total cross sections for methanol over an extensive range of impact
  energy (0.1 -- 2000 eV)}},\ }\href
  {https://doi.org/10.1103/PhysRevA.87.012702} {\bibfield  {journal} {\bibinfo
  {journal} {Phys. Rev. A}\ }\textbf {\bibinfo {volume} {87}},\ \bibinfo
  {pages} {012702} (\bibinfo {year} {2013})}\BibitemShut {NoStop}%
\bibitem [{\citenamefont {Lee}\ \emph {et~al.}(2012)\citenamefont {Lee},
  \citenamefont {de~Souza}, \citenamefont {Machado}, \citenamefont
  {Brescansin}, \citenamefont {dos Santos}, \citenamefont {Lucchese},
  \citenamefont {Sugohara}, \citenamefont {Homem}, \citenamefont {Sanches},\
  and\ \citenamefont {Iga}}]{Lee:2012}%
  \BibitemOpen
  \bibfield  {author} {\bibinfo {author} {\bibfnamefont {M.-T.}\ \bibnamefont
  {Lee}}, \bibinfo {author} {\bibfnamefont {G.~L.~C.}\ \bibnamefont
  {de~Souza}}, \bibinfo {author} {\bibfnamefont {L.~E.}\ \bibnamefont
  {Machado}}, \bibinfo {author} {\bibfnamefont {L.~M.}\ \bibnamefont
  {Brescansin}}, \bibinfo {author} {\bibfnamefont {A.~S.}\ \bibnamefont {dos
  Santos}}, \bibinfo {author} {\bibfnamefont {R.~R.}\ \bibnamefont {Lucchese}},
  \bibinfo {author} {\bibfnamefont {R.~T.}\ \bibnamefont {Sugohara}}, \bibinfo
  {author} {\bibfnamefont {M.~G.~P.}\ \bibnamefont {Homem}}, \bibinfo {author}
  {\bibfnamefont {I.~P.}\ \bibnamefont {Sanches}},\ and\ \bibinfo {author}
  {\bibfnamefont {I.}~\bibnamefont {Iga}},\ }\bibfield  {title} {\bibinfo
  {title} {{Electron scattering by methanol and ethanol: A joint
  theoretical-experimental investigation}},\ }\href
  {https://doi.org/10.1063/1.3695211} {\bibfield  {journal} {\bibinfo
  {journal} {J. Chem. Phys.}\ }\textbf {\bibinfo {volume} {136}},\ \bibinfo
  {pages} {114311} (\bibinfo {year} {2012})}\BibitemShut {NoStop}%
\bibitem [{\citenamefont {Bouchiha}\ \emph {et~al.}(2007)\citenamefont
  {Bouchiha}, \citenamefont {Gorfinkiel}, \citenamefont {Caron},\ and\
  \citenamefont {Sanche}}]{Bouchiha_2007}%
  \BibitemOpen
  \bibfield  {author} {\bibinfo {author} {\bibfnamefont {D.}~\bibnamefont
  {Bouchiha}}, \bibinfo {author} {\bibfnamefont {J.~D.}\ \bibnamefont
  {Gorfinkiel}}, \bibinfo {author} {\bibfnamefont {L.~G.}\ \bibnamefont
  {Caron}},\ and\ \bibinfo {author} {\bibfnamefont {L.}~\bibnamefont
  {Sanche}},\ }\bibfield  {title} {\bibinfo {title} {{Low-energy electron
  collisions with methanol}},\ }\href
  {https://doi.org/10.1088/0953-4075/40/6/016} {\bibfield  {journal} {\bibinfo
  {journal} {J. Phys. B: At. Mol. Opt. Phys.}\ }\textbf {\bibinfo {volume}
  {40}},\ \bibinfo {pages} {1259} (\bibinfo {year} {2007})}\BibitemShut
  {NoStop}%
\bibitem [{\citenamefont {Khakoo}\ \emph
  {et~al.}(2008{\natexlab{a}})\citenamefont {Khakoo}, \citenamefont {Blumer},
  \citenamefont {Keane}, \citenamefont {Campbell}, \citenamefont {Silva},
  \citenamefont {Lopes}, \citenamefont {Winstead}, \citenamefont {McKoy},
  \citenamefont {da~Costa}, \citenamefont {Ferreira}, \citenamefont {Lima},\
  and\ \citenamefont {Bettega}}]{Khakoo_methanol:2008}%
  \BibitemOpen
  \bibfield  {author} {\bibinfo {author} {\bibfnamefont {M.~A.}\ \bibnamefont
  {Khakoo}}, \bibinfo {author} {\bibfnamefont {J.}~\bibnamefont {Blumer}},
  \bibinfo {author} {\bibfnamefont {K.}~\bibnamefont {Keane}}, \bibinfo
  {author} {\bibfnamefont {C.}~\bibnamefont {Campbell}}, \bibinfo {author}
  {\bibfnamefont {H.}~\bibnamefont {Silva}}, \bibinfo {author} {\bibfnamefont
  {M.~C.~A.}\ \bibnamefont {Lopes}}, \bibinfo {author} {\bibfnamefont
  {C.}~\bibnamefont {Winstead}}, \bibinfo {author} {\bibfnamefont
  {V.}~\bibnamefont {McKoy}}, \bibinfo {author} {\bibfnamefont {R.~F.}\
  \bibnamefont {da~Costa}}, \bibinfo {author} {\bibfnamefont {L.~G.}\
  \bibnamefont {Ferreira}}, \bibinfo {author} {\bibfnamefont {M.~A.~P.}\
  \bibnamefont {Lima}},\ and\ \bibinfo {author} {\bibfnamefont {M.~H.~F.}\
  \bibnamefont {Bettega}},\ }\bibfield  {title} {\bibinfo {title} {{Low-energy
  electron scattering from methanol and ethanol}},\ }\href
  {https://doi.org/10.1103/PhysRevA.77.042705} {\bibfield  {journal} {\bibinfo
  {journal} {Phys. Rev. A}\ }\textbf {\bibinfo {volume} {77}},\ \bibinfo
  {pages} {042705} (\bibinfo {year} {2008}{\natexlab{a}})}\BibitemShut
  {NoStop}%
\bibitem [{\citenamefont {Sugohara}\ \emph {et~al.}(2011)\citenamefont
  {Sugohara}, \citenamefont {Homem}, \citenamefont {Sanches}, \citenamefont
  {de~Moura}, \citenamefont {Lee},\ and\ \citenamefont {Iga}}]{Sugohara:2011}%
  \BibitemOpen
  \bibfield  {author} {\bibinfo {author} {\bibfnamefont {R.~T.}\ \bibnamefont
  {Sugohara}}, \bibinfo {author} {\bibfnamefont {M.~G.~P.}\ \bibnamefont
  {Homem}}, \bibinfo {author} {\bibfnamefont {I.~P.}\ \bibnamefont {Sanches}},
  \bibinfo {author} {\bibfnamefont {A.~F.}\ \bibnamefont {de~Moura}}, \bibinfo
  {author} {\bibfnamefont {M.-T.}\ \bibnamefont {Lee}},\ and\ \bibinfo {author}
  {\bibfnamefont {I.}~\bibnamefont {Iga}},\ }\bibfield  {title} {\bibinfo
  {title} {{Cross sections for electron scattering by methanol in the
  intermediate-energy range}},\ }\href
  {https://doi.org/10.1103/PhysRevA.83.032708} {\bibfield  {journal} {\bibinfo
  {journal} {Phys. Rev. A}\ }\textbf {\bibinfo {volume} {83}},\ \bibinfo
  {pages} {032708} (\bibinfo {year} {2011})}\BibitemShut {NoStop}%
\bibitem [{\citenamefont {Brunger}(2017)}]{Brunger_2017}%
  \BibitemOpen
  \bibfield  {author} {\bibinfo {author} {\bibfnamefont {M.~J.}\ \bibnamefont
  {Brunger}},\ }\bibfield  {title} {\bibinfo {title} {{Electron scattering and
  transport in biofuels, biomolecules and biomass fragments}},\ }\href
  {https://doi.org/10.1080/0144235X.2017.1301030} {\bibfield  {journal}
  {\bibinfo  {journal} {Int. Rev. Phys. Chem.}\ }\textbf {\bibinfo {volume}
  {36}},\ \bibinfo {pages} {333} (\bibinfo {year} {2017})}\BibitemShut
  {NoStop}%
\bibitem [{\citenamefont {{\DJ}uri{\'c}}\ \emph {et~al.}(1989)\citenamefont
  {{\DJ}uri{\'c}}, \citenamefont {{\v{C}}ade{\v{z}}},\ and\ \citenamefont
  {Kurepa}}]{djuric:1990}%
  \BibitemOpen
  \bibfield  {author} {\bibinfo {author} {\bibfnamefont {N.}~\bibnamefont
  {{\DJ}uri{\'c}}}, \bibinfo {author} {\bibfnamefont {I.}~\bibnamefont
  {{\v{C}}ade{\v{z}}}},\ and\ \bibinfo {author} {\bibfnamefont
  {M.}~\bibnamefont {Kurepa}},\ }\bibfield  {title} {\bibinfo {title} {{Total
  electron impact ionization cross sections of Methanol, Ethanol and n-Propanol
  molecules}},\ }\href {https://hrcak.srce.hr/331725} {\bibfield  {journal}
  {\bibinfo  {journal} {Fizika A}\ }\textbf {\bibinfo {volume} {21}},\ \bibinfo
  {pages} {339} (\bibinfo {year} {1989})}\BibitemShut {NoStop}%
\bibitem [{\citenamefont {Rejoub}\ \emph {et~al.}(2003)\citenamefont {Rejoub},
  \citenamefont {Morton}, \citenamefont {Lindsay},\ and\ \citenamefont
  {Stebbings}}]{Rejoub:2003}%
  \BibitemOpen
  \bibfield  {author} {\bibinfo {author} {\bibfnamefont {R.}~\bibnamefont
  {Rejoub}}, \bibinfo {author} {\bibfnamefont {C.~D.}\ \bibnamefont {Morton}},
  \bibinfo {author} {\bibfnamefont {B.~G.}\ \bibnamefont {Lindsay}},\ and\
  \bibinfo {author} {\bibfnamefont {R.~F.}\ \bibnamefont {Stebbings}},\
  }\bibfield  {title} {\bibinfo {title} {{Electron-impact ionization of the
  simple alcohols}},\ }\href {https://doi.org/10.1063/1.1531631} {\bibfield
  {journal} {\bibinfo  {journal} {J. Chem. Phys.}\ }\textbf {\bibinfo {volume}
  {118}},\ \bibinfo {pages} {1756} (\bibinfo {year} {2003})}\BibitemShut
  {NoStop}%
\bibitem [{\citenamefont {Srivastava}\ \emph {et~al.}(1996)\citenamefont
  {Srivastava}, \citenamefont {Krishnakumar}, \citenamefont {Fucaloro},\ and\
  \citenamefont {van Note}}]{Srivastava:1996}%
  \BibitemOpen
  \bibfield  {author} {\bibinfo {author} {\bibfnamefont {S.~K.}\ \bibnamefont
  {Srivastava}}, \bibinfo {author} {\bibfnamefont {E.}~\bibnamefont
  {Krishnakumar}}, \bibinfo {author} {\bibfnamefont {A.~F.}\ \bibnamefont
  {Fucaloro}},\ and\ \bibinfo {author} {\bibfnamefont {T.}~\bibnamefont {van
  Note}},\ }\bibfield  {title} {\bibinfo {title} {{Cross sections for the
  production of cations by electron impact on methanol}},\ }\href
  {https://doi.org/https://doi.org/10.1029/96JE02471} {\bibfield  {journal}
  {\bibinfo  {journal} {J. Geophys. Res.: Planets}\ }\textbf {\bibinfo {volume}
  {101}},\ \bibinfo {pages} {26155} (\bibinfo {year} {1996})}\BibitemShut
  {NoStop}%
\bibitem [{\citenamefont {Hudson}\ \emph {et~al.}(2003)\citenamefont {Hudson},
  \citenamefont {Hamilton}, \citenamefont {Vallance},\ and\ \citenamefont
  {Harland}}]{Hudson:2003}%
  \BibitemOpen
  \bibfield  {author} {\bibinfo {author} {\bibfnamefont {J.~E.}\ \bibnamefont
  {Hudson}}, \bibinfo {author} {\bibfnamefont {M.~L.}\ \bibnamefont
  {Hamilton}}, \bibinfo {author} {\bibfnamefont {C.}~\bibnamefont {Vallance}},\
  and\ \bibinfo {author} {\bibfnamefont {P.~W.}\ \bibnamefont {Harland}},\
  }\bibfield  {title} {\bibinfo {title} {{Absolute electron impact ionization
  cross-sections for the C$_1$ to C$_4$ alcohols}},\ }\href
  {https://doi.org/10.1039/B304456D} {\bibfield  {journal} {\bibinfo  {journal}
  {Phys. Chem. Chem. Phys.}\ }\textbf {\bibinfo {volume} {5}},\ \bibinfo
  {pages} {3162} (\bibinfo {year} {2003})}\BibitemShut {NoStop}%
\bibitem [{\citenamefont {Deutsch}\ \emph {et~al.}(1998)\citenamefont
  {Deutsch}, \citenamefont {Becker}, \citenamefont {Basner}, \citenamefont
  {Schmidt},\ and\ \citenamefont {M{\"a}rk}}]{deutsch:1998}%
  \BibitemOpen
  \bibfield  {author} {\bibinfo {author} {\bibfnamefont {H.}~\bibnamefont
  {Deutsch}}, \bibinfo {author} {\bibfnamefont {K.}~\bibnamefont {Becker}},
  \bibinfo {author} {\bibfnamefont {R.}~\bibnamefont {Basner}}, \bibinfo
  {author} {\bibfnamefont {M.}~\bibnamefont {Schmidt}},\ and\ \bibinfo {author}
  {\bibfnamefont {T.}~\bibnamefont {M{\"a}rk}},\ }\bibfield  {title} {\bibinfo
  {title} {{Application of the modified additivity rule to the calculation of
  electron-impact ionization cross sections of complex molecules}},\ }\href
  {https://doi.org/10.1021/jp9827577} {\bibfield  {journal} {\bibinfo
  {journal} {J. Phys. Chem. A}\ }\textbf {\bibinfo {volume} {102}},\ \bibinfo
  {pages} {8819} (\bibinfo {year} {1998})}\BibitemShut {NoStop}%
\bibitem [{\citenamefont {Pal}(2004)}]{PAL:2004}%
  \BibitemOpen
  \bibfield  {author} {\bibinfo {author} {\bibfnamefont {S.}~\bibnamefont
  {Pal}},\ }\bibfield  {title} {\bibinfo {title} {{Determination of single
  differential and partial cross-sections for the production of cations in
  electron–methanol collision}},\ }\href
  {https://doi.org/https://doi.org/10.1016/j.chemphys.2004.03.029} {\bibfield
  {journal} {\bibinfo  {journal} {Chem. Phys.}\ }\textbf {\bibinfo {volume}
  {302}},\ \bibinfo {pages} {119} (\bibinfo {year} {2004})}\BibitemShut
  {NoStop}%
\bibitem [{\citenamefont {Vinodkumar}\ \emph {et~al.}(2011)\citenamefont
  {Vinodkumar}, \citenamefont {Korot},\ and\ \citenamefont
  {Vinodkumar}}]{VINODKUMAR:2011}%
  \BibitemOpen
  \bibfield  {author} {\bibinfo {author} {\bibfnamefont {M.}~\bibnamefont
  {Vinodkumar}}, \bibinfo {author} {\bibfnamefont {K.}~\bibnamefont {Korot}},\
  and\ \bibinfo {author} {\bibfnamefont {P.}~\bibnamefont {Vinodkumar}},\
  }\bibfield  {title} {\bibinfo {title} {{Computation of the electron impact
  total ionization cross sections of C$_n$H$_{(2n+1)}$OH molecules from the
  threshold to 2keV energy range}},\ }\href
  {https://doi.org/https://doi.org/10.1016/j.ijms.2011.04.005} {\bibfield
  {journal} {\bibinfo  {journal} {Int. J. Mass Spectrom.}\ }\textbf {\bibinfo
  {volume} {305}},\ \bibinfo {pages} {26} (\bibinfo {year} {2011})}\BibitemShut
  {NoStop}%
\bibitem [{\citenamefont {Trepka}\ and\ \citenamefont
  {Neuert}(1963)}]{trepka:1963}%
  \BibitemOpen
  \bibfield  {author} {\bibinfo {author} {\bibfnamefont {L.}~\bibnamefont
  {Trepka}}\ and\ \bibinfo {author} {\bibfnamefont {H.}~\bibnamefont
  {Neuert}},\ }\bibfield  {title} {\bibinfo {title} {{{\"U}ber die Entstehung
  von negativen Ionen aus einigen Kohlenwasserstoffen und Alkoholen durch
  Elektronensto{\ss}}},\ }\href {https://doi.org/10.1515/zna-1963-1208}
  {\bibfield  {journal} {\bibinfo  {journal} {Z. Naturforsch. A}\ }\textbf
  {\bibinfo {volume} {18}},\ \bibinfo {pages} {1295} (\bibinfo {year}
  {1963})}\BibitemShut {NoStop}%
\bibitem [{\citenamefont {Kühn}\ \emph {et~al.}(1988)\citenamefont {Kühn},
  \citenamefont {Fenzlaff},\ and\ \citenamefont {Illenberger}}]{Kuhn:1988}%
  \BibitemOpen
  \bibfield  {author} {\bibinfo {author} {\bibfnamefont {A.}~\bibnamefont
  {Kühn}}, \bibinfo {author} {\bibfnamefont {H.}~\bibnamefont {Fenzlaff}},\
  and\ \bibinfo {author} {\bibfnamefont {E.}~\bibnamefont {Illenberger}},\
  }\bibfield  {title} {\bibinfo {title} {{Formation and dissociation of
  negative ion resonances in methanol and allylalcohol}},\ }\href
  {https://doi.org/10.1063/1.454309} {\bibfield  {journal} {\bibinfo  {journal}
  {J. Chem. Phys.}\ }\textbf {\bibinfo {volume} {88}},\ \bibinfo {pages} {7453}
  (\bibinfo {year} {1988})}\BibitemShut {NoStop}%
\bibitem [{\citenamefont {Curtis}\ and\ \citenamefont
  {Walker}(1992)}]{Curtis:1992}%
  \BibitemOpen
  \bibfield  {author} {\bibinfo {author} {\bibfnamefont {M.~G.}\ \bibnamefont
  {Curtis}}\ and\ \bibinfo {author} {\bibfnamefont {I.~C.}\ \bibnamefont
  {Walker}},\ }\bibfield  {title} {\bibinfo {title} {{Dissociative electron
  attachment in water and methanol (5–14 eV)}},\ }\href
  {https://doi.org/10.1039/FT9928802805} {\bibfield  {journal} {\bibinfo
  {journal} {J. Chem. Soc.{,} Faraday Trans.}\ }\textbf {\bibinfo {volume}
  {88}},\ \bibinfo {pages} {2805} (\bibinfo {year} {1992})}\BibitemShut
  {NoStop}%
\bibitem [{\citenamefont {Skalický}\ and\ \citenamefont
  {Allan}(2004)}]{Skalicky_2004}%
  \BibitemOpen
  \bibfield  {author} {\bibinfo {author} {\bibfnamefont {T.}~\bibnamefont
  {Skalický}}\ and\ \bibinfo {author} {\bibfnamefont {M.}~\bibnamefont
  {Allan}},\ }\bibfield  {title} {\bibinfo {title} {{The assignment of
  dissociative electron attachment bands in compounds containing hydroxyl and
  amino groups}},\ }\href {https://doi.org/10.1088/0953-4075/37/24/010}
  {\bibfield  {journal} {\bibinfo  {journal} {J. Phys. B: At. Mol. Opt. Phys.}\
  }\textbf {\bibinfo {volume} {37}},\ \bibinfo {pages} {4849} (\bibinfo {year}
  {2004})}\BibitemShut {NoStop}%
\bibitem [{\citenamefont {Prabhudesai}\ \emph {et~al.}(2008)\citenamefont
  {Prabhudesai}, \citenamefont {Nandi}, \citenamefont {Kelkar},\ and\
  \citenamefont {Krishnakumar}}]{Prabhudesai_2008_jcp}%
  \BibitemOpen
  \bibfield  {author} {\bibinfo {author} {\bibfnamefont {V.~S.}\ \bibnamefont
  {Prabhudesai}}, \bibinfo {author} {\bibfnamefont {D.}~\bibnamefont {Nandi}},
  \bibinfo {author} {\bibfnamefont {A.~H.}\ \bibnamefont {Kelkar}},\ and\
  \bibinfo {author} {\bibfnamefont {E.}~\bibnamefont {Krishnakumar}},\
  }\bibfield  {title} {\bibinfo {title} {{Functional group dependent
  dissociative electron attachment to simple organic molecules}},\ }\href
  {https://doi.org/10.1063/1.2899330} {\bibfield  {journal} {\bibinfo
  {journal} {J. Chem. Phys.}\ }\textbf {\bibinfo {volume} {128}},\ \bibinfo
  {pages} {154309} (\bibinfo {year} {2008})}\BibitemShut {NoStop}%
\bibitem [{\citenamefont {Wang}\ \emph {et~al.}(2015)\citenamefont {Wang},
  \citenamefont {Xuan}, \citenamefont {Feng},\ and\ \citenamefont
  {Tian}}]{Wang:2015}%
  \BibitemOpen
  \bibfield  {author} {\bibinfo {author} {\bibfnamefont {X.-D.}\ \bibnamefont
  {Wang}}, \bibinfo {author} {\bibfnamefont {C.-J.}\ \bibnamefont {Xuan}},
  \bibinfo {author} {\bibfnamefont {W.-L.}\ \bibnamefont {Feng}},\ and\
  \bibinfo {author} {\bibfnamefont {S.~X.}\ \bibnamefont {Tian}},\ }\bibfield
  {title} {\bibinfo {title} {{Dissociative electron attachments to ethanol and
  acetaldehyde: A combined experimental and simulation study}},\ }\href
  {https://doi.org/10.1063/1.4907940} {\bibfield  {journal} {\bibinfo
  {journal} {J. Chem. Phys.}\ }\textbf {\bibinfo {volume} {142}},\ \bibinfo
  {pages} {064316} (\bibinfo {year} {2015})}\BibitemShut {NoStop}%
\bibitem [{\citenamefont {Orzol}\ \emph {et~al.}(2007)\citenamefont {Orzol},
  \citenamefont {Martin}, \citenamefont {Kocisek}, \citenamefont {Dabkowska},
  \citenamefont {Langer},\ and\ \citenamefont {Illenberger}}]{orzol:2007}%
  \BibitemOpen
  \bibfield  {author} {\bibinfo {author} {\bibfnamefont {M.}~\bibnamefont
  {Orzol}}, \bibinfo {author} {\bibfnamefont {I.}~\bibnamefont {Martin}},
  \bibinfo {author} {\bibfnamefont {J.}~\bibnamefont {Kocisek}}, \bibinfo
  {author} {\bibfnamefont {I.}~\bibnamefont {Dabkowska}}, \bibinfo {author}
  {\bibfnamefont {J.}~\bibnamefont {Langer}},\ and\ \bibinfo {author}
  {\bibfnamefont {E.}~\bibnamefont {Illenberger}},\ }\bibfield  {title}
  {\bibinfo {title} {{Bond and site selectivity in dissociative electron
  attachment to gas phase and condensed phase ethanol and trifluoroethanol}},\
  }\href {https://doi.org/10.1039/B701543G} {\bibfield  {journal} {\bibinfo
  {journal} {Phys. Chem. Chem. Phys.}\ }\textbf {\bibinfo {volume} {9}},\
  \bibinfo {pages} {3424} (\bibinfo {year} {2007})}\BibitemShut {NoStop}%
\bibitem [{\citenamefont {Khakoo}\ \emph
  {et~al.}(2008{\natexlab{b}})\citenamefont {Khakoo}, \citenamefont {Muse},
  \citenamefont {Silva}, \citenamefont {Lopes}, \citenamefont {Winstead},
  \citenamefont {McKoy}, \citenamefont {de~Oliveira}, \citenamefont {da~Costa},
  \citenamefont {Varella}, \citenamefont {Bettega},\ and\ \citenamefont
  {Lima}}]{Khakoo:2008}%
  \BibitemOpen
  \bibfield  {author} {\bibinfo {author} {\bibfnamefont {M.~A.}\ \bibnamefont
  {Khakoo}}, \bibinfo {author} {\bibfnamefont {J.}~\bibnamefont {Muse}},
  \bibinfo {author} {\bibfnamefont {H.}~\bibnamefont {Silva}}, \bibinfo
  {author} {\bibfnamefont {M.~C.~A.}\ \bibnamefont {Lopes}}, \bibinfo {author}
  {\bibfnamefont {C.}~\bibnamefont {Winstead}}, \bibinfo {author}
  {\bibfnamefont {V.}~\bibnamefont {McKoy}}, \bibinfo {author} {\bibfnamefont
  {E.~M.}\ \bibnamefont {de~Oliveira}}, \bibinfo {author} {\bibfnamefont
  {R.~F.}\ \bibnamefont {da~Costa}}, \bibinfo {author} {\bibfnamefont {M.~T.
  d.~N.}\ \bibnamefont {Varella}}, \bibinfo {author} {\bibfnamefont {M.~H.~F.}\
  \bibnamefont {Bettega}},\ and\ \bibinfo {author} {\bibfnamefont {M.~A.~P.}\
  \bibnamefont {Lima}},\ }\bibfield  {title} {\bibinfo {title} {{Elastic
  scattering of slow electrons by $n$-propanol and $n$-butanol}},\ }\href
  {https://doi.org/10.1103/PhysRevA.78.062714} {\bibfield  {journal} {\bibinfo
  {journal} {Phys. Rev. A}\ }\textbf {\bibinfo {volume} {78}},\ \bibinfo
  {pages} {062714} (\bibinfo {year} {2008}{\natexlab{b}})}\BibitemShut
  {NoStop}%
\bibitem [{\citenamefont {Williams}\ and\ \citenamefont
  {Hamill}(1968)}]{Williams:1968}%
  \BibitemOpen
  \bibfield  {author} {\bibinfo {author} {\bibfnamefont {J.~M.}\ \bibnamefont
  {Williams}}\ and\ \bibinfo {author} {\bibfnamefont {W.~H.}\ \bibnamefont
  {Hamill}},\ }\bibfield  {title} {\bibinfo {title} {{Ionization Potentials of
  Molecules and Free Radicals and Appearance Potentials by Electron Impact in
  the Mass Spectrometer}},\ }\href {https://doi.org/10.1063/1.1669899}
  {\bibfield  {journal} {\bibinfo  {journal} {J. Chem. Phys.}\ }\textbf
  {\bibinfo {volume} {49}},\ \bibinfo {pages} {4467} (\bibinfo {year}
  {1968})}\BibitemShut {NoStop}%
\bibitem [{\citenamefont {Bull}\ \emph {et~al.}(2012)\citenamefont {Bull},
  \citenamefont {Harland},\ and\ \citenamefont {Vallance}}]{Bull:2012}%
  \BibitemOpen
  \bibfield  {author} {\bibinfo {author} {\bibfnamefont {J.~N.}\ \bibnamefont
  {Bull}}, \bibinfo {author} {\bibfnamefont {P.~W.}\ \bibnamefont {Harland}},\
  and\ \bibinfo {author} {\bibfnamefont {C.}~\bibnamefont {Vallance}},\
  }\bibfield  {title} {\bibinfo {title} {{Absolute Total Electron Impact
  Ionization Cross-Sections for Many-Atom Organic and Halocarbon Species}},\
  }\href {https://doi.org/10.1021/jp210294p} {\bibfield  {journal} {\bibinfo
  {journal} {J. Phys. Chem. A}\ }\textbf {\bibinfo {volume} {116}},\ \bibinfo
  {pages} {767} (\bibinfo {year} {2012})}\BibitemShut {NoStop}%
\bibitem [{\citenamefont {Takeuchi}\ \emph {et~al.}(1985)\citenamefont
  {Takeuchi}, \citenamefont {Ueno}, \citenamefont {Yamamoto}, \citenamefont
  {Matsushita},\ and\ \citenamefont {Nishimoto}}]{TAKEUCHI:1985}%
  \BibitemOpen
  \bibfield  {author} {\bibinfo {author} {\bibfnamefont {T.}~\bibnamefont
  {Takeuchi}}, \bibinfo {author} {\bibfnamefont {S.}~\bibnamefont {Ueno}},
  \bibinfo {author} {\bibfnamefont {M.}~\bibnamefont {Yamamoto}}, \bibinfo
  {author} {\bibfnamefont {T.}~\bibnamefont {Matsushita}},\ and\ \bibinfo
  {author} {\bibfnamefont {K.}~\bibnamefont {Nishimoto}},\ }\bibfield  {title}
  {\bibinfo {title} {{Theoretical study of electron impact mass spectrometry.
  II. ab initio MO study of the fragmentation of ionized 1-propanol}},\ }\href
  {https://doi.org/https://doi.org/10.1016/0168-1176(85)85034-5} {\bibfield
  {journal} {\bibinfo  {journal} {Int. J. Mass Spectrom. Ion Process.}\
  }\textbf {\bibinfo {volume} {64}},\ \bibinfo {pages} {33} (\bibinfo {year}
  {1985})}\BibitemShut {NoStop}%
\bibitem [{\citenamefont {Pires}\ \emph {et~al.}(2017)\citenamefont {Pires},
  \citenamefont {Nixon}, \citenamefont {Ghosh}, \citenamefont {Neves},
  \citenamefont {Duque}, \citenamefont {Amorim}, \citenamefont {Jones},
  \citenamefont {Blanco}, \citenamefont {Garcia}, \citenamefont {Brunger},\
  and\ \citenamefont {Lopes}}]{PIRES:2017}%
  \BibitemOpen
  \bibfield  {author} {\bibinfo {author} {\bibfnamefont {W.}~\bibnamefont
  {Pires}}, \bibinfo {author} {\bibfnamefont {K.}~\bibnamefont {Nixon}},
  \bibinfo {author} {\bibfnamefont {S.}~\bibnamefont {Ghosh}}, \bibinfo
  {author} {\bibfnamefont {R.}~\bibnamefont {Neves}}, \bibinfo {author}
  {\bibfnamefont {H.}~\bibnamefont {Duque}}, \bibinfo {author} {\bibfnamefont
  {R.}~\bibnamefont {Amorim}}, \bibinfo {author} {\bibfnamefont
  {D.}~\bibnamefont {Jones}}, \bibinfo {author} {\bibfnamefont
  {F.}~\bibnamefont {Blanco}}, \bibinfo {author} {\bibfnamefont
  {G.}~\bibnamefont {Garcia}}, \bibinfo {author} {\bibfnamefont
  {M.}~\bibnamefont {Brunger}},\ and\ \bibinfo {author} {\bibfnamefont
  {M.}~\bibnamefont {Lopes}},\ }\bibfield  {title} {\bibinfo {title} {{Electron
  impact ionization of 1-propanol}},\ }\href
  {https://doi.org/https://doi.org/10.1016/j.ijms.2017.08.005} {\bibfield
  {journal} {\bibinfo  {journal} {Int. J. Mass Spectrom.}\ }\textbf {\bibinfo
  {volume} {422}},\ \bibinfo {pages} {32} (\bibinfo {year} {2017})}\BibitemShut
  {NoStop}%
\bibitem [{\citenamefont {da~Silva}\ \emph {et~al.}(2017)\citenamefont
  {da~Silva}, \citenamefont {Gomes}, \citenamefont {Ghosh}, \citenamefont
  {Silva}, \citenamefont {Pires}, \citenamefont {Jones}, \citenamefont
  {Blanco}, \citenamefont {Garcia}, \citenamefont {Buckman}, \citenamefont
  {Brunger},\ and\ \citenamefont {Lopes}}]{Silva:2017}%
  \BibitemOpen
  \bibfield  {author} {\bibinfo {author} {\bibfnamefont {D.~G.~M.}\
  \bibnamefont {da~Silva}}, \bibinfo {author} {\bibfnamefont {M.}~\bibnamefont
  {Gomes}}, \bibinfo {author} {\bibfnamefont {S.}~\bibnamefont {Ghosh}},
  \bibinfo {author} {\bibfnamefont {I.~F.~L.}\ \bibnamefont {Silva}}, \bibinfo
  {author} {\bibfnamefont {W.~A.~D.}\ \bibnamefont {Pires}}, \bibinfo {author}
  {\bibfnamefont {D.~B.}\ \bibnamefont {Jones}}, \bibinfo {author}
  {\bibfnamefont {F.}~\bibnamefont {Blanco}}, \bibinfo {author} {\bibfnamefont
  {G.}~\bibnamefont {Garcia}}, \bibinfo {author} {\bibfnamefont {S.~J.}\
  \bibnamefont {Buckman}}, \bibinfo {author} {\bibfnamefont {M.~J.}\
  \bibnamefont {Brunger}},\ and\ \bibinfo {author} {\bibfnamefont {M.~C.~A.}\
  \bibnamefont {Lopes}},\ }\bibfield  {title} {\bibinfo {title} {{Total cross
  sections for electron scattering by 1-propanol at impact energies in the
  range 40-500 eV}},\ }\href {https://doi.org/10.1063/1.5008621} {\bibfield
  {journal} {\bibinfo  {journal} {J. Chem. Phys.}\ }\textbf {\bibinfo {volume}
  {147}},\ \bibinfo {pages} {194307} (\bibinfo {year} {2017})}\BibitemShut
  {NoStop}%
\bibitem [{\citenamefont {Ibănescu}(2009)}]{IbanescuBC:phd}%
  \BibitemOpen
  \bibfield  {author} {\bibinfo {author} {\bibfnamefont {B.~C.}\ \bibnamefont
  {Ibănescu}},\ }\emph {\bibinfo {title} {Electron-Driven Chemistry of
  Saturated Compounds Containing Oxygen or Nitrogen Atoms}},\ \href
  {https://sonar.rero.ch/documents/301679/files/IbanescuBC.pdf} {Ph.D.
  thesis},\ \bibinfo  {school} {University of Freiburg} (\bibinfo {year}
  {2009})\BibitemShut {NoStop}%
\bibitem [{\citenamefont {Chakraborty}\ \emph {et~al.}(2018)\citenamefont
  {Chakraborty}, \citenamefont {Nag},\ and\ \citenamefont
  {Nandi}}]{Chakraborty:2018}%
  \BibitemOpen
  \bibfield  {author} {\bibinfo {author} {\bibfnamefont {D.}~\bibnamefont
  {Chakraborty}}, \bibinfo {author} {\bibfnamefont {P.}~\bibnamefont {Nag}},\
  and\ \bibinfo {author} {\bibfnamefont {D.}~\bibnamefont {Nandi}},\ }\bibfield
   {title} {\bibinfo {title} {{A new time of flight mass spectrometer for
  absolute dissociative electron attachment cross-section measurements in gas
  phase}},\ }\href {https://doi.org/10.1063/1.5017656} {\bibfield  {journal}
  {\bibinfo  {journal} {Rev. Sci. Instrum.}\ }\textbf {\bibinfo {volume}
  {89}},\ \bibinfo {pages} {025115} (\bibinfo {year} {2018})}\BibitemShut
  {NoStop}%
\bibitem [{\citenamefont {Rapp}\ and\ \citenamefont
  {Briglia}(2004)}]{Rapp:2004}%
  \BibitemOpen
  \bibfield  {author} {\bibinfo {author} {\bibfnamefont {D.}~\bibnamefont
  {Rapp}}\ and\ \bibinfo {author} {\bibfnamefont {D.~D.}\ \bibnamefont
  {Briglia}},\ }\bibfield  {title} {\bibinfo {title} {{Total Cross Sections for
  Ionization and Attachment in Gases by Electron Impact. II. Negative‐Ion
  Formation}},\ }\href {https://doi.org/10.1063/1.1696958} {\bibfield
  {journal} {\bibinfo  {journal} {J. Chem. Phys.}\ }\textbf {\bibinfo {volume}
  {43}},\ \bibinfo {pages} {1480} (\bibinfo {year} {2004})}\BibitemShut
  {NoStop}%
\bibitem [{\citenamefont {Frisch}\ \emph {et~al.}(2016)\citenamefont {Frisch},
  \citenamefont {Trucks}, \citenamefont {Schlegel}, \citenamefont {Scuseria},
  \citenamefont {Robb}, \citenamefont {Cheeseman}, \citenamefont {Scalmani},
  \citenamefont {Barone}, \citenamefont {Petersson}, \citenamefont {Nakatsuji},
  \citenamefont {Li}, \citenamefont {Caricato}, \citenamefont {Marenich},
  \citenamefont {Bloino}, \citenamefont {Janesko}, \citenamefont {Gomperts},
  \citenamefont {Mennucci}, \citenamefont {Hratchian}, \citenamefont {Ortiz},
  \citenamefont {Izmaylov}, \citenamefont {Sonnenberg}, \citenamefont
  {Williams-Young}, \citenamefont {Ding}, \citenamefont {Lipparini},
  \citenamefont {Egidi}, \citenamefont {Goings}, \citenamefont {Peng},
  \citenamefont {Petrone}, \citenamefont {Henderson}, \citenamefont
  {Ranasinghe}, \citenamefont {Zakrzewski}, \citenamefont {Gao}, \citenamefont
  {Rega}, \citenamefont {Zheng}, \citenamefont {Liang}, \citenamefont {Hada},
  \citenamefont {Ehara}, \citenamefont {Toyota}, \citenamefont {Fukuda},
  \citenamefont {Hasegawa}, \citenamefont {Ishida}, \citenamefont {Nakajima},
  \citenamefont {Honda}, \citenamefont {Kitao}, \citenamefont {Nakai},
  \citenamefont {Vreven}, \citenamefont {Throssell}, \citenamefont
  {Montgomery}, \citenamefont {Peralta}, \citenamefont {Ogliaro}, \citenamefont
  {Bearpark}, \citenamefont {Heyd}, \citenamefont {Brothers}, \citenamefont
  {Kudin}, \citenamefont {Staroverov}, \citenamefont {Keith}, \citenamefont
  {Kobayashi}, \citenamefont {Normand}, \citenamefont {Raghavachari},
  \citenamefont {Rendell}, \citenamefont {Burant}, \citenamefont {Iyengar},
  \citenamefont {Tomasi}, \citenamefont {Cossi}, \citenamefont {Millam},
  \citenamefont {Klene}, \citenamefont {Adamo}, \citenamefont {Cammi},
  \citenamefont {Ochterski}, \citenamefont {Martin}, \citenamefont {Morokuma},
  \citenamefont {Farkas}, \citenamefont {Foresman},\ and\ \citenamefont
  {Fox}}]{g16}%
  \BibitemOpen
  \bibfield  {author} {\bibinfo {author} {\bibfnamefont {M.~J.}\ \bibnamefont
  {Frisch}}, \bibinfo {author} {\bibfnamefont {G.~W.}\ \bibnamefont {Trucks}},
  \bibinfo {author} {\bibfnamefont {H.~B.}\ \bibnamefont {Schlegel}}, \bibinfo
  {author} {\bibfnamefont {G.~E.}\ \bibnamefont {Scuseria}}, \bibinfo {author}
  {\bibfnamefont {M.~A.}\ \bibnamefont {Robb}}, \bibinfo {author}
  {\bibfnamefont {J.~R.}\ \bibnamefont {Cheeseman}}, \bibinfo {author}
  {\bibfnamefont {G.}~\bibnamefont {Scalmani}}, \bibinfo {author}
  {\bibfnamefont {V.}~\bibnamefont {Barone}}, \bibinfo {author} {\bibfnamefont
  {G.~A.}\ \bibnamefont {Petersson}}, \bibinfo {author} {\bibfnamefont
  {H.}~\bibnamefont {Nakatsuji}}, \bibinfo {author} {\bibfnamefont
  {X.}~\bibnamefont {Li}}, \bibinfo {author} {\bibfnamefont {M.}~\bibnamefont
  {Caricato}}, \bibinfo {author} {\bibfnamefont {A.~V.}\ \bibnamefont
  {Marenich}}, \bibinfo {author} {\bibfnamefont {J.}~\bibnamefont {Bloino}},
  \bibinfo {author} {\bibfnamefont {B.~G.}\ \bibnamefont {Janesko}}, \bibinfo
  {author} {\bibfnamefont {R.}~\bibnamefont {Gomperts}}, \bibinfo {author}
  {\bibfnamefont {B.}~\bibnamefont {Mennucci}}, \bibinfo {author}
  {\bibfnamefont {H.~P.}\ \bibnamefont {Hratchian}}, \bibinfo {author}
  {\bibfnamefont {J.~V.}\ \bibnamefont {Ortiz}}, \bibinfo {author}
  {\bibfnamefont {A.~F.}\ \bibnamefont {Izmaylov}}, \bibinfo {author}
  {\bibfnamefont {J.~L.}\ \bibnamefont {Sonnenberg}}, \bibinfo {author}
  {\bibfnamefont {D.}~\bibnamefont {Williams-Young}}, \bibinfo {author}
  {\bibfnamefont {F.}~\bibnamefont {Ding}}, \bibinfo {author} {\bibfnamefont
  {F.}~\bibnamefont {Lipparini}}, \bibinfo {author} {\bibfnamefont
  {F.}~\bibnamefont {Egidi}}, \bibinfo {author} {\bibfnamefont
  {J.}~\bibnamefont {Goings}}, \bibinfo {author} {\bibfnamefont
  {B.}~\bibnamefont {Peng}}, \bibinfo {author} {\bibfnamefont {A.}~\bibnamefont
  {Petrone}}, \bibinfo {author} {\bibfnamefont {T.}~\bibnamefont {Henderson}},
  \bibinfo {author} {\bibfnamefont {D.}~\bibnamefont {Ranasinghe}}, \bibinfo
  {author} {\bibfnamefont {V.~G.}\ \bibnamefont {Zakrzewski}}, \bibinfo
  {author} {\bibfnamefont {J.}~\bibnamefont {Gao}}, \bibinfo {author}
  {\bibfnamefont {N.}~\bibnamefont {Rega}}, \bibinfo {author} {\bibfnamefont
  {G.}~\bibnamefont {Zheng}}, \bibinfo {author} {\bibfnamefont
  {W.}~\bibnamefont {Liang}}, \bibinfo {author} {\bibfnamefont
  {M.}~\bibnamefont {Hada}}, \bibinfo {author} {\bibfnamefont {M.}~\bibnamefont
  {Ehara}}, \bibinfo {author} {\bibfnamefont {K.}~\bibnamefont {Toyota}},
  \bibinfo {author} {\bibfnamefont {R.}~\bibnamefont {Fukuda}}, \bibinfo
  {author} {\bibfnamefont {J.}~\bibnamefont {Hasegawa}}, \bibinfo {author}
  {\bibfnamefont {M.}~\bibnamefont {Ishida}}, \bibinfo {author} {\bibfnamefont
  {T.}~\bibnamefont {Nakajima}}, \bibinfo {author} {\bibfnamefont
  {Y.}~\bibnamefont {Honda}}, \bibinfo {author} {\bibfnamefont
  {O.}~\bibnamefont {Kitao}}, \bibinfo {author} {\bibfnamefont
  {H.}~\bibnamefont {Nakai}}, \bibinfo {author} {\bibfnamefont
  {T.}~\bibnamefont {Vreven}}, \bibinfo {author} {\bibfnamefont
  {K.}~\bibnamefont {Throssell}}, \bibinfo {author} {\bibfnamefont {J.~A.}\
  \bibnamefont {Montgomery}, \bibfnamefont {{Jr.}}}, \bibinfo {author}
  {\bibfnamefont {J.~E.}\ \bibnamefont {Peralta}}, \bibinfo {author}
  {\bibfnamefont {F.}~\bibnamefont {Ogliaro}}, \bibinfo {author} {\bibfnamefont
  {M.~J.}\ \bibnamefont {Bearpark}}, \bibinfo {author} {\bibfnamefont {J.~J.}\
  \bibnamefont {Heyd}}, \bibinfo {author} {\bibfnamefont {E.~N.}\ \bibnamefont
  {Brothers}}, \bibinfo {author} {\bibfnamefont {K.~N.}\ \bibnamefont {Kudin}},
  \bibinfo {author} {\bibfnamefont {V.~N.}\ \bibnamefont {Staroverov}},
  \bibinfo {author} {\bibfnamefont {T.~A.}\ \bibnamefont {Keith}}, \bibinfo
  {author} {\bibfnamefont {R.}~\bibnamefont {Kobayashi}}, \bibinfo {author}
  {\bibfnamefont {J.}~\bibnamefont {Normand}}, \bibinfo {author} {\bibfnamefont
  {K.}~\bibnamefont {Raghavachari}}, \bibinfo {author} {\bibfnamefont {A.~P.}\
  \bibnamefont {Rendell}}, \bibinfo {author} {\bibfnamefont {J.~C.}\
  \bibnamefont {Burant}}, \bibinfo {author} {\bibfnamefont {S.~S.}\
  \bibnamefont {Iyengar}}, \bibinfo {author} {\bibfnamefont {J.}~\bibnamefont
  {Tomasi}}, \bibinfo {author} {\bibfnamefont {M.}~\bibnamefont {Cossi}},
  \bibinfo {author} {\bibfnamefont {J.~M.}\ \bibnamefont {Millam}}, \bibinfo
  {author} {\bibfnamefont {M.}~\bibnamefont {Klene}}, \bibinfo {author}
  {\bibfnamefont {C.}~\bibnamefont {Adamo}}, \bibinfo {author} {\bibfnamefont
  {R.}~\bibnamefont {Cammi}}, \bibinfo {author} {\bibfnamefont {J.~W.}\
  \bibnamefont {Ochterski}}, \bibinfo {author} {\bibfnamefont {R.~L.}\
  \bibnamefont {Martin}}, \bibinfo {author} {\bibfnamefont {K.}~\bibnamefont
  {Morokuma}}, \bibinfo {author} {\bibfnamefont {O.}~\bibnamefont {Farkas}},
  \bibinfo {author} {\bibfnamefont {J.~B.}\ \bibnamefont {Foresman}},\ and\
  \bibinfo {author} {\bibfnamefont {D.~J.}\ \bibnamefont {Fox}},\ }\href@noop
  {} {\bibinfo {title} {{Gaussian~16 {R}evision {C}.01}}} (\bibinfo {year}
  {2016}),\ \bibinfo {note} {gaussian Inc. Wallingford CT}\BibitemShut
  {NoStop}%
\bibitem [{\citenamefont {Becke}(1993)}]{Becke_1993}%
  \BibitemOpen
  \bibfield  {author} {\bibinfo {author} {\bibfnamefont {A.~D.}\ \bibnamefont
  {Becke}},\ }\bibfield  {title} {\bibinfo {title} {{Density‐functional
  thermochemistry. III. The role of exact exchange}},\ }\href
  {https://doi.org/10.1063/1.464913} {\bibfield  {journal} {\bibinfo  {journal}
  {J. Chem. Phys.}\ }\textbf {\bibinfo {volume} {98}},\ \bibinfo {pages} {5648}
  (\bibinfo {year} {1993})}\BibitemShut {NoStop}%
\bibitem [{\citenamefont {Dunning}(1989)}]{Dunning:1989}%
  \BibitemOpen
  \bibfield  {author} {\bibinfo {author} {\bibfnamefont {J.}~\bibnamefont
  {Dunning}, \bibfnamefont {Thom~H.}},\ }\bibfield  {title} {\bibinfo {title}
  {{Gaussian basis sets for use in correlated molecular calculations. I. The
  atoms boron through neon and hydrogen}},\ }\href
  {https://doi.org/10.1063/1.456153} {\bibfield  {journal} {\bibinfo  {journal}
  {J. Chem. Phys.}\ }\textbf {\bibinfo {volume} {90}},\ \bibinfo {pages} {1007}
  (\bibinfo {year} {1989})}\BibitemShut {NoStop}%
\bibitem [{\citenamefont {Ibănescu}\ and\ \citenamefont
  {Allan}(2009)}]{Ibanesu_Ethanol}%
  \BibitemOpen
  \bibfield  {author} {\bibinfo {author} {\bibfnamefont {B.~C.}\ \bibnamefont
  {Ibănescu}}\ and\ \bibinfo {author} {\bibfnamefont {M.}~\bibnamefont
  {Allan}},\ }\bibfield  {title} {\bibinfo {title} {{Selective cleavage of the
  C–O bonds in alcohols and asymmetric ethers by dissociative electron
  attachment}},\ }\href {https://doi.org/10.1039/B904945B} {\bibfield
  {journal} {\bibinfo  {journal} {Phys. Chem. Chem. Phys.}\ }\textbf {\bibinfo
  {volume} {11}},\ \bibinfo {pages} {7640} (\bibinfo {year}
  {2009})}\BibitemShut {NoStop}%
\bibitem [{\citenamefont {Fedor}\ \emph {et~al.}(2006)\citenamefont {Fedor},
  \citenamefont {Cicman}, \citenamefont {Coupier}, \citenamefont {Feil},
  \citenamefont {Winkler}, \citenamefont {Głuch}, \citenamefont {Husarik},
  \citenamefont {Jaksch}, \citenamefont {Farizon}, \citenamefont {Mason},
  \citenamefont {Scheier},\ and\ \citenamefont {Märk}}]{Fedor_2006}%
  \BibitemOpen
  \bibfield  {author} {\bibinfo {author} {\bibfnamefont {J.}~\bibnamefont
  {Fedor}}, \bibinfo {author} {\bibfnamefont {P.}~\bibnamefont {Cicman}},
  \bibinfo {author} {\bibfnamefont {B.}~\bibnamefont {Coupier}}, \bibinfo
  {author} {\bibfnamefont {S.}~\bibnamefont {Feil}}, \bibinfo {author}
  {\bibfnamefont {M.}~\bibnamefont {Winkler}}, \bibinfo {author} {\bibfnamefont
  {K.}~\bibnamefont {Głuch}}, \bibinfo {author} {\bibfnamefont
  {J.}~\bibnamefont {Husarik}}, \bibinfo {author} {\bibfnamefont
  {D.}~\bibnamefont {Jaksch}}, \bibinfo {author} {\bibfnamefont
  {B.}~\bibnamefont {Farizon}}, \bibinfo {author} {\bibfnamefont {N.~J.}\
  \bibnamefont {Mason}}, \bibinfo {author} {\bibfnamefont {P.}~\bibnamefont
  {Scheier}},\ and\ \bibinfo {author} {\bibfnamefont {T.~D.}\ \bibnamefont
  {Märk}},\ }\bibfield  {title} {\bibinfo {title} {{Fragmentation of transient
  water anions following low-energy electron capture by H$_2$O/D$_2$O}},\
  }\href {https://doi.org/10.1088/0953-4075/39/18/022} {\bibfield  {journal}
  {\bibinfo  {journal} {J. Phys. B: At. Mol. Opt. Phys.}\ }\textbf {\bibinfo
  {volume} {39}},\ \bibinfo {pages} {3935} (\bibinfo {year}
  {2006})}\BibitemShut {NoStop}%
\bibitem [{\citenamefont {Rawat}\ \emph {et~al.}(2007)\citenamefont {Rawat},
  \citenamefont {Prabhudesai}, \citenamefont {Aravind}, \citenamefont
  {Bhargavaram}, \citenamefont {Rahman},\ and\ \citenamefont
  {Krishnakumar}}]{Rawat_2007}%
  \BibitemOpen
  \bibfield  {author} {\bibinfo {author} {\bibfnamefont {P.}~\bibnamefont
  {Rawat}}, \bibinfo {author} {\bibfnamefont {V.~S.}\ \bibnamefont
  {Prabhudesai}}, \bibinfo {author} {\bibfnamefont {G.}~\bibnamefont
  {Aravind}}, \bibinfo {author} {\bibfnamefont {N.}~\bibnamefont
  {Bhargavaram}}, \bibinfo {author} {\bibfnamefont {M.~A.}\ \bibnamefont
  {Rahman}},\ and\ \bibinfo {author} {\bibfnamefont {E.}~\bibnamefont
  {Krishnakumar}},\ }\bibfield  {title} {\bibinfo {title} {{Absolute cross
  sections for dissociative electron attachment to water, methane and
  ammonia}},\ }\href {https://doi.org/10.1088/1742-6596/80/1/012018} {\bibfield
   {journal} {\bibinfo  {journal} {J. Phys.: Conf. Ser.}\ }\textbf {\bibinfo
  {volume} {80}},\ \bibinfo {pages} {012018} (\bibinfo {year}
  {2007})}\BibitemShut {NoStop}%
\end{thebibliography}%

\end{document}